\documentclass[english,preprint]{aastex}
\usepackage[T1]{fontenc}
\usepackage[latin9]{inputenc}
\usepackage{amsmath}
\usepackage{amssymb}



\usepackage{babel}

\begin{document}

\title{Dispersion of Magnetic Fields in Molecular Clouds. II}

\author{Martin Houde$^{1}$, John E. Vaillancourt$^{2}$, Roger H. Hildebrand$^{3,4}$,
Shadi Chitsazzadeh$^{1}$, and Larry Kirby$^{3}$ }

\affil{$^{1}$Department of Physics and Astronomy, The University of Western
Ontario, London, ON, N6A 3K7, Canada}

\affil{$^{2}$Division of Physics, Mathematics, \& Astronomy, California
Institute of Technology, Pasadena, CA 91125}

\affil{$^{3}$Department of Astronomy and Astrophysics and Enrico Fermi
Institute, The University of Chicago, Chicago, IL 60637}

\affil{$^{4}$Department of Physics, The University of Chicago, Chicago,
IL 60637}
\begin{abstract}
We expand our study on the dispersion of polarization angles in molecular
clouds. We show how the effect of signal integration through the thickness
of the cloud as well as across the area subtended by the telescope
beam inherent to dust continuum measurements can be incorporated in
our analysis to correctly account for its effect on the measured angular
dispersion and inferred turbulent to large-scale magnetic field strength
ratio. We further show how to evaluate the turbulent magnetic field
correlation scale from polarization data of sufficient spatial resolution
and high enough spatial sampling rate. We apply our results to the
molecular cloud OMC-1, where we find a turbulent correlation length
of $\delta\approx16$ mpc, a turbulent to large-scale magnetic field
strength ratio of approximately 0.5, and a plane-of-the-sky large-scale
magnetic field strength of approximately $760\,\mu\mathrm{G}$.
\end{abstract}


\section{Introduction}

The observational determination of the turbulent energy content within
the magnetic field is important for understanding the role of magnetic
fields in the star formation process. It provides, for example, some
measure of the amount of turbulent energy contained in the gas while
the further determination of the turbulent to large-scale magnetic
field strength ratio can in principle be used to evaluate the large-scale
magnetic strength with the so-called Chandrasekhar-Fermi equation
\citep{CF1953}. 

The determination of the turbulent to large-scale magnetic field strength
ratio was the subject of a recent publication (\citealt{Hildebrand2009};
hereafter Paper I) where it was shown how this parameter can be precisely
extracted from dust continuum polarization data through a careful
analysis of polarization angle differences as a function of the distance
between pair of points where measurements were made (i.e., the angular
dispersion function). It was described how the evaluation of the turbulent
polarization angular dispersion can be achieved without assuming any
model for the large-scale magnetic field component about which this
dispersion is calculated. This is an important development since no
model will perfectly fit the true morphology of the large-scale magnetic
field. Fits to a model will therefore lead to inaccurate estimates
of the angular dispersion. This error would then be propagated in
the determination, for example, of the large-scale magnetic field
strength when the Chandrasekhar-Fermi equation is used.

In this second paper on the subject we generalize our analysis of
Paper I by including the process of signal integration through the
thickness of the cloud as well as across the area subtended by the
telescope beam inherent to dust continuum measurements. It has long
been recognized that the amount of angular dispersion measured in
a polarization map is reduced by any integration process \citep{Myers1991},
as has since been studied and demonstrated through numerical simulations
\citep{Ostriker2001, Padoan2001, Heitsch2001, Kudoh2003, Falceta2008}.
The effect of integration through the thickness of a cloud has been
considered by \citet{Myers1991} in their studies of the optical polarization
of dark clouds, while the further inclusion of integration across
the area subtended by the telescope beam and ensuing consequences
on measurements were investigated through simulations by \citet{Heitsch2001},
\citet{Wiebe2004}, and \citet{Falceta2008}.

We will start with a generalization of the problem considered in Paper
I by deriving the cloud- and beam-integrated dispersion function in
Section \ref{sec:Analysis}, which will then be solved for the special
case of Gaussian turbulent autocorrelation and beam profile functions.
In Section \ref{sec:Results} we apply our analysis to dust continuum
polarization data obtained with SHARP \citep{Novak2004, Li2006, Li2008a}
for the molecular cloud OMC-1. We show how the turbulent correlation
length scale for this cloud can be evaluated with the corresponding
polarization data. We then use these results to determine the number
of independent turbulent cells contained in the column of dust probed
with our measurements and calculate the turbulent to large-scale magnetic
field strength ratio corrected for the signal integration process.
We provide a detailed discussion of our results in Section \ref{sec:Discussion}
and end with a summary in Section \ref{sec:Summary}. Detailed derivations
resulting in the relations and functions used in these sections, description
of the data analysis, as well as a list of variables and functions
will be found in the appendices at the end of the paper.

\section{Analysis\label{sec:Analysis}}

\subsection{The Cloud- and Beam-integrated Angular Dispersion Function}

In Paper I the following equation was introduced (see their Equations
{[}A4{]} and {[}A20{]}) for the analytical derivation of the dispersion
in polarization angles within a turbulent molecular cloud

\begin{equation}
\left\langle \cos\left[\Delta\Phi\left(\boldsymbol{\ell}\right)\right]\right\rangle =\frac{\left\langle \mathbf{B\left(\mathbf{x}\right)\cdot}\mathbf{B\left(\mathbf{x+\boldsymbol{\ell}}\right)}\right\rangle }{\left[\left\langle B^{2}\left(\mathbf{x}\right)\right\rangle \left\langle B^{2}\left(\mathbf{x}+\boldsymbol{\ell}\right)\right\rangle \right]^{1/2}},\label{eq:cos_noint}\end{equation}

\noindent where $\Delta\Phi\left(\boldsymbol{\ell}\right)\equiv\Phi\left(\mathbf{x}\right)-\Phi\left(\mathbf{x}+\boldsymbol{\ell}\right)$
is the difference in the polarization angle $\Phi$ measured at two
positions separated by a distance $\boldsymbol{\ell}$ and $\left\langle \cdots\right\rangle $
denotes an average. As was then shown, Equation (\ref{eq:cos_noint})
(and others that derive from it) applies equally well to a three-dimensional
magnetic field or its two-dimensional projection onto a plane; for
the purpose of this paper we are considering $\mathbf{B\left(\mathbf{x}\right)}$
to be the plane-of-the-sky projected magnetic field, as usually probed
with dust continuum polarization measurements. The distance $\boldsymbol{\ell}$
is also confined to the plane-of-the-sky, unless otherwise noted.
We assume that the magnetic field $\mathbf{B\left(\mathbf{x}\right)}$
is composed of a large-scale, structured field, $\mathbf{B}_{0}\mathbf{\left(\mathbf{x}\right)}$,
and a turbulent (or random) component, $\mathbf{B_{\mathrm{t}}\left(\mathbf{x}\right)}$,
such that 

\begin{equation}
\mathbf{\mathbf{B\left(\mathbf{x}\right)}}=\mathbf{B}_{0}\mathbf{\mathbf{\mathbf{\left(\mathbf{x}\right)+\mathbf{B}_{\mathrm{t}}\left(\mathbf{x}\right)}}}.\label{eq:Btot}\end{equation}

We must note, however, that the magnetic field direction inferred
from polarization data is actually the result of some averaging process
as one integrates through the thickness of the cloud along the line-of-sight
as well as across the area subtended by the telescope beam. We therefore
first define the normalized magnetic field vector 

\begin{equation}
\mathbf{b}\left(\mathbf{r},z\right)\equiv\frac{\mathbf{B}\left(\mathbf{r},z\right)}{\left\langle B^{2}\left(\mathbf{r},z\right)\right\rangle ^{1/2}}\label{eq:b(r,z)}\end{equation}

\noindent and infer a mean direction for the cloud-integrated magnetic
field through the following weighted integral of $\mathbf{b}\left(\mathbf{r},z\right)$

\begin{equation}
\overline{\mathbf{b}}\left(\mathbf{r}\right)\equiv\iint H\left(\mathbf{r}-\mathbf{a}\right)\left[\frac{1}{\Delta}\int_{0}^{\Delta}F\left(\mathbf{a},z\right)\mathbf{b}\left(\mathbf{a},z\right)dz\right]d^{2}a.\label{eq:bbar}\end{equation}

\noindent The two-dimensional (convolution) integral in Equation (\ref{eq:bbar})
is over all space such that for the three-dimensional position vector
$\mathbf{x}$, $\mathbf{r}$ is the two-dimensional polar radius vector
on the plane-of-the-sky and $z$ the depth within the cloud. That
is, 

\begin{equation}
\mathbf{x}=r\mathbf{e}_{r}+z\mathbf{e}_{z}\label{eq:x}\end{equation}

\noindent with $\mathbf{e}_{r}$ and $\mathbf{e}_{z}$ the unit basis
vectors along $\mathbf{r}$ and the $z$-axis (which is oriented along
the line-of-sight), respectively. The beam profile density is denoted
by $H\left(\mathbf{r}\right)$, while the weighting function $F\left(\mathbf{r},z\right)\geq0$
is the polarized emission associated with the magnetic field $\mathbf{b}\left(\mathbf{r},z\right)$%
\footnote{The measured linear polarization orientation is normal to that of
the associated plane-of-the-sky magnetic field when detecting dust
continuum emission (at submillimeter wavelengths, for example), not
parallel to it as could be inferred from Equation (\ref{eq:bbar}).
This is irrelevant to our analysis, however, as we are using the polarized
emission as a weighting function in defining a mean orientation for
the integrated magnetic field. %
}. Please note that even though $\mathbf{b}\left(\mathbf{r},z\right)$
is normalized $\overline{\mathbf{b}}\left(\mathbf{r}\right)$ is not,
and while the normalization by $\Delta$ in Equation (\ref{eq:bbar})
is not essential it is included for convenience. Moreover, the quantity
$\Delta$ is for the maximum depth of the cloud along any line-of-sight;
the detailed behavior of the $F\left(\mathbf{r},z\right)$ function
thus ensures that Equation (\ref{eq:bbar}) is exact even when the
depth of the cloud is expected to vary with position on the plane-of-the-sky.

The normalization of the magnetic field vector through the cloud is
warranted because the amount of polarized emission in a given region
is not a function of the strength of the magnetic field itself. Because
of this we must now consider a slightly different relation for determining
the dispersion function. That is, we replace Equation (\ref{eq:cos_noint})
with 

\begin{equation}
\left\langle \cos\left[\Delta\Phi\left(\boldsymbol{\ell}\right)\right]\right\rangle \equiv\frac{\left\langle \overline{\mathbf{b}}\mathbf{\left(\mathbf{r}\right)\cdot}\overline{\mathbf{b}}\left(\mathbf{r+\boldsymbol{\ell}}\right)\right\rangle }{\left[\left\langle \overline{b}^{2}\left(\mathbf{r}\right)\right\rangle \left\langle \overline{b}^{2}\left(\mathbf{r}+\boldsymbol{\ell}\right)\right\rangle \right]^{1/2}}.\label{eq:cos_int}\end{equation}

In what follows we introduce a level of idealization necessary for
obtaining a quantitative measure of the turbulent component of the
magnetic field in molecular clouds. We assume stationarity, homogeneity
and isotropy in the magnetic field strength, as well as statistical
independence between its large-scale and turbulent components. We
therefore have the following averages at points $\mathbf{x}$ and
$\mathbf{y}$

\begin{eqnarray}
\left\langle \mathbf{B}_{0}\left(\mathbf{x}\right)\right\rangle  & = & \mathbf{B}_{0}\left(\mathbf{x}\right)\nonumber \\
\left\langle \mathbf{B}_{\mathrm{t}}\left(\mathbf{x}\right)\right\rangle  & = & 0\nonumber \\
\left\langle \mathbf{B}_{0}\left(\mathbf{x}\right)\cdot\mathbf{B}_{\mathrm{t}}\left(\mathbf{y}\right)\right\rangle  & = & \left\langle \mathbf{B}_{0}\left(\mathbf{x}\right)\right\rangle \cdot\left\langle \mathbf{B}_{\mathrm{t}}\left(\mathbf{y}\right)\right\rangle =0,\label{eq:averages}\end{eqnarray}

\noindent and

\begin{eqnarray}
\left\langle \mathbf{B}_{0}^{2}\left(\mathbf{x}\right)\right\rangle  & = & \left\langle \mathbf{B}_{0}^{2}\left(\mathbf{y}\right)\right\rangle =\left\langle B_{0}^{2}\right\rangle \nonumber \\
\left\langle \mathbf{B}_{\mathrm{t}}^{2}\left(\mathbf{x}\right)\right\rangle  & = & \left\langle \mathbf{B}_{\mathrm{t}}^{2}\left(\mathbf{y}\right)\right\rangle =\left\langle B_{\mathrm{t}}^{2}\right\rangle .\label{eq:homo}\end{eqnarray}

It is straightforward to show that the homogeneity in the field strength
renders the field normalization of Equation (\ref{eq:b(r,z)}) inconsequential
in Equation (\ref{eq:cos_int}) and we can therefore equally write

\begin{equation}
\left\langle \cos\left[\Delta\Phi\left(\ell\right)\right]\right\rangle =\frac{\left\langle \overline{\mathbf{B}}\mathbf{\left(\mathbf{r}\right)\cdot}\overline{\mathbf{B}}\mathbf{\left(\mathbf{r+\boldsymbol{\ell}}\right)}\right\rangle }{\left[\left\langle \overline{B}^{2}\left(\mathbf{r}\right)\right\rangle \left\langle \overline{B}^{2}\left(\mathbf{r}+\boldsymbol{\ell}\right)\right\rangle \right]^{1/2}},\label{eq:cos_final}\end{equation}

\noindent with the cloud- and beam-integrated magnetic field

\begin{equation}
\overline{\mathbf{B}}\left(\mathbf{r}\right)=\iint H\left(\mathbf{r}-\mathbf{a}\right)\left[\frac{1}{\Delta}\int_{0}^{\Delta}F\left(\mathbf{a},z\right)\mathbf{B}\left(\mathbf{a},z\right)dz\right]d^{2}a\label{eq:Bbar}\end{equation}

\noindent and where the assumed isotropy in the distance ($\ell=\left|\boldsymbol{\ell}\right|$)
was incorporated. Equation (\ref{eq:cos_final}) is the relation we
will use to estimate the angular dispersion function.

\subsection{The Integrated Magnetic Field Autocorrelation Function }

In view of the assumed stationarity and isotropy the integrated magnetic
field autocorrelation function $\left\langle \overline{\mathbf{B}}\mathbf{\cdot}\overline{\mathbf{B}}\mathbf{\left(\ell\right)}\right\rangle \equiv\left\langle \overline{\mathbf{B}}\mathbf{\left(\mathbf{r}\right)\cdot}\overline{\mathbf{B}}\mathbf{\left(\mathbf{r+\boldsymbol{\ell}}\right)}\right\rangle $
can be expressed as (see Eqs. {[}\ref{eq:autoBbar_a}{]}-{[}\ref{eq:autoBbar_app}{]}
in Appendix \ref{sec:Derivations})

\begin{equation}
\left\langle \overline{\mathbf{B}}\mathbf{\cdot}\overline{\mathbf{B}}\mathbf{\left(\ell\right)}\right\rangle =\iint\iint H\left(\mathbf{a}\right)H\left(\mathbf{a}^{\prime}+\boldsymbol{\ell}\right)\left[\frac{2}{\Delta}\int_{0}^{\Delta}\left(1-\frac{u}{\Delta}\right)R_{F}\left(v,u\right)R_{\mathbf{B}}\left(v,u\right)du\right]d^{2}a^{\prime}d^{2}a,\label{eq:auto_Bbar}\end{equation}

\noindent where we introduced the (assumed statistically independent)
autocorrelation functions for the (non-integrated) magnetic field
and the polarized emission 

\begin{eqnarray}
R_{\mathbf{B}}\left(v,u\right) & = & \left\langle \mathbf{B}\left(\mathbf{a},z\right)\cdot\mathbf{B}\left(\mathbf{a}^{\prime},z^{\prime}\right)\right\rangle \label{eq:autoB}\\
R_{F}\left(v,u\right) & = & \left\langle F\left(\mathbf{a},z\right)F\left(\mathbf{a}^{\prime},z^{\prime}\right)\right\rangle \label{eq:auto_g}\end{eqnarray}

\noindent with $u=\left|z^{\prime}-z\right|$ and $v=\left|\mathbf{a}^{\prime}-\mathbf{a}\right|$.
These autocorrelations can be further broken down through the decomposition
of the magnetic field and the polarized emission into their respective
large-scale and turbulent components (see Equation {[}\ref{eq:Btot}{]})

\begin{eqnarray}
\mathbf{B}\left(\mathbf{a},z\right) & = & \mathbf{B}_{0}\left(\mathbf{a},z\right)+\mathbf{B}_{\mathrm{t}}\left(\mathbf{a},z\right)\label{eq:B_decomp}\\
F\left(\mathbf{a},z\right) & = & F_{0}\left(\mathbf{a},z\right)+F_{\mathrm{t}}\left(\mathbf{a},z\right).\label{eq:g_decomp}\end{eqnarray}

\noindent Upon assuming the same statistical properties for the components
of $F\left(\mathbf{a},z\right)$ as those expressed in Equations (\ref{eq:averages})
for $\mathbf{B}\left(\mathbf{r},z\right)$, Equations (\ref{eq:autoB})
and (\ref{eq:auto_g}) transform to

\begin{eqnarray}
R_{\mathbf{B}}\left(v,u\right) & = & R_{\mathbf{B},0}\left(v,u\right)+R_{\mathbf{B},\mathrm{t}}\left(v,u\right)\label{eq:autoB_decomp}\\
R_{F}\left(v,u\right) & = & R_{F,0}\left(v,u\right)+R_{F,\mathrm{t}}\left(v,u\right)\label{eq:autog_decomp}\end{eqnarray}

\noindent with

\begin{eqnarray}
R_{\mathbf{B},j}\left(v,u\right) & = & \left\langle \mathbf{B}_{j}\left(\mathbf{a},z\right)\cdot\mathbf{B}_{j}\left(\mathbf{a}^{\prime},z^{\prime}\right)\right\rangle \label{eq:autoB_j}\\
R_{F,j}\left(v,u\right) & = & \left\langle F_{j}\left(\mathbf{a},z\right)F_{j}\left(\mathbf{a}^{\prime},z^{\prime}\right)\right\rangle \label{eq:autog_j}\end{eqnarray}

\noindent where $j=\,$'0' or 't' for the large-scale and turbulent
components, respectively.

The solution to our problem is reduced to solving Equation (\ref{eq:auto_Bbar})
given the different autocorrelation functions; the equation to be
used for determining the dispersion function then becomes

\begin{equation}
\left\langle \cos\left[\Delta\Phi\left(\ell\right)\right]\right\rangle =\frac{\left\langle \overline{\mathbf{B}}\mathbf{\cdot}\overline{\mathbf{B}}\mathbf{\left(\ell\right)}\right\rangle }{\left\langle \overline{\mathbf{B}}\mathbf{\cdot}\overline{\mathbf{B}}\left(0\right)\right\rangle }.\label{eq:cos_final2}\end{equation}

\subsection{Solution Using Gaussian Turbulent Autocorrelation and Beam Profile
Functions}

In order to solve Equation (\ref{eq:cos_final2}) we must specify,
at least to some level, the characteristics of the different autocorrelation
functions, as well as the telescope beam profile. We therefore define
the following

\begin{eqnarray}
R_{\mathbf{B}}\left(v,u\right) & = & R_{\mathbf{B},0}\left(v,u\right)+\left\langle B_{\mathrm{t}}^{2}\right\rangle e^{-\left(v^{2}+u^{2}\right)/2\delta^{2}}\label{eq:autoB_gen}\\
R_{F}\left(v,u\right) & = & R_{F,0}\left(v,u\right)+\left\langle F_{\mathrm{t}}^{2}\right\rangle e^{-\left(v^{2}+u^{2}\right)/2\delta^{\prime2}},\label{eq:autog_gen}\end{eqnarray}

\noindent where the magnetic field and polarized emission autocorrelations
each have large-scale and turbulent (second term on the right-hand
sides) components. The correlation length scales for the turbulent
magnetic field and polarized emission are $\delta$ and $\delta^{\prime}$,
respectively, and, as is implied through Equations (\ref{eq:autoB_gen})
and (\ref{eq:autog_gen}), we assume the turbulence to be isotropic.
Furthermore, $\delta$ and $\delta^{\prime}$ are taken to be much
smaller than the thickness of the cloud (i.e., $\delta\ll\Delta$
and $\delta^{\prime}\ll\Delta$). The beam profile is defined with

\begin{equation}
H\left(\mathbf{r}\right)=\frac{1}{2\pi W^{2}}e^{-r^{2}/2W^{2}},\label{eq:beam}\end{equation}

\noindent where $W$ is the beam {}``radius''.

Under these constraints and definitions Equation (\ref{eq:auto_Bbar})
can be analytically solved to yield (see Eqs. {[}\ref{eq:onedim_integ}{]}-{[}\ref{eq:beam_integ}{]}
Appendix \ref{sec:Derivations})

\begin{eqnarray}
\left\langle \overline{\mathbf{B}}\mathbf{\cdot}\overline{\mathbf{B}}\mathbf{\left(\ell\right)}\right\rangle  & \simeq & \left\langle B_{0}^{2}\right\rangle \left\langle F_{0}^{2}\right\rangle \left(\left\langle \alpha\left(\ell\right)\right\rangle +\sqrt{2\pi}\frac{\left\langle B_{\mathrm{t}}^{2}\right\rangle }{\left\langle B_{0}^{2}\right\rangle }\left\{ \left[\frac{\delta^{3}}{\left(\delta^{2}+2W^{2}\right)\Delta}\right]e^{-\ell^{2}/2\left(\delta^{2}+2W^{2}\right)}\right.\right.\nonumber \\
 &  & \left.+\frac{\left\langle F_{\mathrm{t}}^{2}\right\rangle }{\left\langle F_{0}^{2}\right\rangle }\left[\frac{\delta^{\prime\prime3}}{\left(\delta^{\prime\prime2}+2W^{2}\right)\Delta}\right]e^{-\ell^{2}/2\left(\delta^{\prime\prime2}+2W^{2}\right)}\right\} \nonumber \\
 &  & \left.+\sqrt{2\pi}\frac{\left\langle F_{\mathrm{t}}^{2}\right\rangle }{\left\langle F_{0}^{2}\right\rangle }\left[\frac{\delta^{\prime3}}{\left(\delta^{\prime2}+2W^{2}\right)\Delta}\right]e^{-\ell^{2}/2\left(\delta^{\prime2}+2W^{2}\right)}\right),\label{eq:auto_Bbar_sol}\end{eqnarray}

\noindent with

\begin{eqnarray}
\left\langle B_{0}^{2}\right\rangle  & = & R_{\mathbf{B},0}\left(0,0\right)\label{eq:B_0^2}\\
\left\langle F_{0}^{2}\right\rangle  & = & R_{F,0}\left(0,0\right)\label{eq:F_0^2}\\
\delta^{\prime\prime} & = & \frac{\delta\delta^{\prime}}{\sqrt{\delta^{2}+\delta^{\prime2}}}\label{eq:delta''}\end{eqnarray}

\noindent and the normalized large-scale function

\begin{equation}
\left\langle \alpha\left(\ell\right)\right\rangle =\iint\iint H\left(\mathbf{a}\right)H\left(\mathbf{a}^{\prime}+\boldsymbol{\ell}\right)\left\{ \frac{2}{\Delta}\int_{0}^{\Delta}\left(1-\frac{u}{\Delta}\right)\left[\frac{R_{F,0}\left(v,u\right)}{\left\langle F_{0}^{2}\right\rangle }\right]\left[\frac{R_{\mathbf{B},0}\left(v,u\right)}{\left\langle B_{0}^{2}\right\rangle }\right]du\right\} d^{2}a^{\prime}d^{2}a.\label{eq:norm_ls}\end{equation}

It is expected that this large-scale function will usually dominate
the other terms in Equation (\ref{eq:auto_Bbar_sol}), since these
all result from the averaging of the turbulent components through
the column of dust probed by the corresponding polarization measurements.
Indeed, it is apparent that each of the turbulent contributions in
Equation (\ref{eq:auto_Bbar_sol}) scales with a term such as

\begin{equation}
N^{-1}=\frac{\sqrt{2\pi}\delta^{3}}{\left(\delta^{2}+2W^{2}\right)\Delta},\label{eq:N}\end{equation}

\noindent where $N$ is nothing more than the number of independent
turbulent cells (for the magnetic field in this case) contained in
the column of dust probed observationally, as could have been intuitively
guessed. We also note that for cases where the telescope beam radius
were to be much smaller than the size of a turbulent cell (e.g., $\delta\gg W$)
we recover

\begin{equation}
N=\frac{\Delta}{\sqrt{2\pi}\delta}.\label{eq:N_1D}\end{equation}
The number of independent turbulent cells would thus be accounted
for by those cells that lie along the line-of-sight through the thickness
of the cloud at a given point on its surface, as would also be expected
intuitively \citep{Myers1991}. 

Inserting Equation (\ref{eq:auto_Bbar_sol}) through (\ref{eq:norm_ls})
into Equation (\ref{eq:cos_final2}) we can write

\begin{eqnarray}
1-\left\langle \cos\left[\Delta\Phi\left(\ell\right)\right]\right\rangle  & = & \frac{\left\langle B_{0}^{2}\right\rangle \left\langle F_{0}^{2}\right\rangle }{\left\langle \overline{\mathbf{B}}\mathbf{\cdot}\overline{\mathbf{B}}\left(0\right)\right\rangle }\left(\rule{0in}{4ex}\left[\left\langle \alpha\left(0\right)\right\rangle -\left\langle \alpha\left(\ell\right)\right\rangle \right]\right.\nonumber \\
 &  & +\sqrt{2\pi}\frac{\left\langle B_{\mathrm{t}}^{2}\right\rangle }{\left\langle B_{0}^{2}\right\rangle }\left\{ \left[\frac{\delta^{3}}{\left(\delta^{2}+2W^{2}\right)\Delta}\right]\left[1-e^{-\ell^{2}/2\left(\delta^{2}+2W^{2}\right)}\right]\right.\nonumber \\
 &  & \left.+\frac{\left\langle F_{\mathrm{t}}^{2}\right\rangle }{\left\langle F_{0}^{2}\right\rangle }\left[\frac{\delta^{\prime\prime3}}{\left(\delta^{\prime\prime2}+2W^{2}\right)\Delta}\right]\left[1-e^{-\ell^{2}/2\left(\delta^{\prime\prime2}+2W^{2}\right)}\right]\right\} \nonumber \\
 &  & \left.+\sqrt{2\pi}\frac{\left\langle F_{\mathrm{t}}^{2}\right\rangle }{\left\langle F_{0}^{2}\right\rangle }\left[\frac{\delta^{\prime3}}{\left(\delta^{\prime2}+2W^{2}\right)\Delta}\right]\left[1-e^{-\ell^{2}/2\left(\delta^{\prime2}+2W^{2}\right)}\right]\right).\label{eq:cos_sol1}\end{eqnarray}

The first term within parentheses on the right-hand side is, as was
previously stated, due to the large-scale structure in the magnetic
field and the polarized emission; it does not involve turbulence.
We can expand this term using a Taylor series with

\begin{equation}
\left\langle \alpha\left(0\right)\right\rangle -\left\langle \alpha\left(\ell\right)\right\rangle =\sum_{j=1}^{\infty}a_{2j}\ell^{2j},\label{eq:taylor_ls}\end{equation}

\noindent where the summation is performed only on even values for
$2j$, since $\left\langle \alpha\left(\ell\right)\right\rangle $
is isotropic in $\ell$ (see Section \ref{sub:ls} in Appendix \ref{sec:Derivations}). 

Although the remaining terms in Equation (\ref{eq:cos_sol1}) are
all due to turbulence, their respective contributions have, in part,
different origins. More importantly, it should be clear that the total
turbulent component is not only due to turbulence in the magnetic
field but can also arise from the presence of turbulence in the polarized
emission. For example, the third term within parentheses on the right-hand
side of Equation (\ref{eq:cos_sol1}) can be interpreted as a contribution
due to random changes in polarized emission at different positions
in the cloud where the magnetic field also changes orientation. It
follows that the measured (i.e., integrated) orientation associated
with the polarized emission, and therefore the deduced magnetic field
orientation (i.e., the polarization angle), will also accordingly
fluctuate randomly.

The last term in Equation (\ref{eq:cos_sol1}) seems to imply that
the turbulent component in polarized emission will contribute to the
turbulent angular dispersion even in cases where the magnetic field
is uniform (and does not have a turbulent component). But this is
an artifact of the way we analytically evaluate the dispersion function
through Equation (\ref{eq:cos_sol1}), which does not perfectly mimic
the way polarization measurements are accomplished (see Section \ref{sec:Discussion}).
In reality, polarization measurements are made on a point by point
basis, and the dispersion function is calculated through an average
of (the cosine of) angle differences as a function of the displacement
$\ell$ (see Eq. {[}1{]} in Paper I) not through the evaluation of
an autocorrelation function such as given in Equation (\ref{eq:auto_Bbar}).
It follows that a perfectly uniform magnetic field could never lead
to a measurable angular dispersion; we will therefore not include
in our analysis the corresponding contribution in Equation (\ref{eq:cos_sol1}). 

Now consider the following

\small

\begin{eqnarray}
\frac{\left\langle \overline{\mathbf{B}}\mathbf{\cdot}\overline{\mathbf{B}}\left(0\right)\right\rangle }{\left\langle B_{0}^{2}\right\rangle \left\langle F_{0}^{2}\right\rangle } & \simeq & \left\langle \alpha\left(0\right)\right\rangle \nonumber \\
 & \simeq & \iint\iint H\left(\mathbf{a}\right)H\left(\mathbf{a}^{\prime}\right)\left\{ \frac{2}{\Delta}\int_{0}^{\Delta}\left(1-\frac{u}{\Delta}\right)\left[\frac{R_{F,0}\left(v,u\right)}{\left\langle F_{0}^{2}\right\rangle }\right]\left[\frac{R_{\mathbf{B},0}\left(v,u\right)}{\left\langle B_{0}^{2}\right\rangle }\right]du\right\} d^{2}a^{\prime}d^{2}a,\label{eq:alpha(0)_app}\end{eqnarray}
\normalsize

\noindent where we have taken advantage of the fact that we expect
that the large-scale component $\left\langle \alpha\left(0\right)\right\rangle $
dominates the turbulent terms in Equation (\ref{eq:auto_Bbar_sol})
(when $\ell=0)$, as was previously stated. Because the autocorrelations
present in the integrand of the one-dimensional integral are normalized
it follows that this integral, which we denote as $A\left(\left|\mathbf{a}^{\prime}-\mathbf{a}\right|\right)/2$,
will always be less than $\Delta/2$ and thus

\begin{eqnarray}
\frac{\left\langle \overline{\mathbf{B}}\mathbf{\cdot}\overline{\mathbf{B}}\left(0\right)\right\rangle }{\left\langle B_{0}^{2}\right\rangle \left\langle F_{0}^{2}\right\rangle } & \simeq & \frac{1}{\Delta}\iint\iint H\left(\mathbf{a}\right)H\left(\mathbf{a}^{\prime}\right)A\left(\left|\mathbf{a}^{\prime}-\mathbf{a}\right|\right)d^{2}a^{\prime}d^{2}a\nonumber \\
 & \equiv & \frac{\Delta^{\prime}}{\Delta}\leq1,\label{eq:Delta'}\end{eqnarray}

\noindent where the equality to unity only holds when the large-scale
magnetic field and polarized emission are both uniform. The quantity
$\Delta^{\prime}$ can be interpreted as the effective thickness of
the cloud; this will be discussed in more details in Section \ref{sub:thickness}.

Taking these considerations into account, Equation (\ref{eq:cos_sol1})
becomes

\begin{eqnarray}
1-\left\langle \cos\left[\Delta\Phi\left(\ell\right)\right]\right\rangle  & \simeq & \sqrt{2\pi}\frac{\left\langle B_{\mathrm{t}}^{2}\right\rangle }{\left\langle B_{0}^{2}\right\rangle }\left\{ \rule{0in}{3.8ex}\left[\frac{\delta^{3}}{\left(\delta^{2}+2W^{2}\right)\Delta^{\prime}}\right]\left[1-e^{-\ell^{2}/2\left(\delta^{2}+2W^{2}\right)}\right]\right.\nonumber \\
 &  & \left.+\frac{\left\langle F_{\mathrm{t}}^{2}\right\rangle }{\left\langle F_{0}^{2}\right\rangle }\left[\frac{\delta^{\prime\prime3}}{\left(\delta^{\prime\prime2}+2W^{2}\right)\Delta^{\prime}}\right]\left[1-e^{-\ell^{2}/2\left(\delta^{\prime\prime2}+2W^{2}\right)}\right]\right\} +\sum_{j=1}^{\infty}a_{2j}^{\prime}\ell^{2j},\label{eq:cos_sol2}\end{eqnarray}

\noindent with $a_{2j}^{\prime}=\left(\Delta/\Delta^{\prime}\right)a_{2j}$.
If we limit ourselves to small enough displacements such that $\ell$
is less than a few times the beam radius $W$, then the large-scale
term becomes small enough to be adequately described by the first
term in the Taylor expansion of Equation (\ref{eq:taylor_ls}) and
Equation (\ref{eq:cos_sol2}) can be approximated to 

\begin{eqnarray}
1-\left\langle \cos\left[\Delta\Phi\left(\ell\right)\right]\right\rangle  & \simeq & \sqrt{2\pi}\frac{\left\langle B_{\mathrm{t}}^{2}\right\rangle }{\left\langle B_{0}^{2}\right\rangle }\left\{ \rule{0in}{3.8ex}\left[\frac{\delta^{3}}{\left(\delta^{2}+2W^{2}\right)\Delta^{\prime}}\right]\left[1-e^{-\ell^{2}/2\left(\delta^{2}+2W^{2}\right)}\right]\right.\nonumber \\
 &  & \left.+\frac{\left\langle F_{\mathrm{t}}^{2}\right\rangle }{\left\langle F_{0}^{2}\right\rangle }\left[\frac{\delta^{\prime\prime3}}{\left(\delta^{\prime\prime2}+2W^{2}\right)\Delta^{\prime}}\right]\left[1-e^{-\ell^{2}/2\left(\delta^{\prime\prime2}+2W^{2}\right)}\right]\right\} +a_{2}^{\prime}\ell^{2},\label{eq:cos_sol2_app}\end{eqnarray}

or alternatively

\begin{eqnarray}
\left\langle \Delta\Phi^{2}\left(\ell\right)\right\rangle  & \simeq & 2\sqrt{2\pi}\frac{\left\langle B_{\mathrm{t}}^{2}\right\rangle }{\left\langle B_{0}^{2}\right\rangle }\left\{ \rule{0in}{3.8ex}\left[\frac{\delta^{3}}{\left(\delta^{2}+2W^{2}\right)\Delta^{\prime}}\right]\left[1-e^{-\ell^{2}/2\left(\delta^{2}+2W^{2}\right)}\right]\right.\nonumber \\
 &  & \left.+\frac{\left\langle F_{\mathrm{t}}^{2}\right\rangle }{\left\langle F_{0}^{2}\right\rangle }\left[\frac{\delta^{\prime\prime3}}{\left(\delta^{\prime\prime2}+2W^{2}\right)\Delta^{\prime}}\right]\left[1-e^{-\ell^{2}/2\left(\delta^{\prime\prime2}+2W^{2}\right)}\right]\right\} +m^{2}\ell^{2},\label{eq:DeltaPhi^2}\end{eqnarray}

where $m^{2}=2a_{2}^{\prime}$.

\section{Results\label{sec:Results}}

\subsection{The Polarized Emission\label{sub:pol}}

Ideally one would intend to use Equation (\ref{eq:cos_sol2_app})
for the dispersion function (or alternatively Equation {[}\ref{eq:DeltaPhi^2}{]})
to determine the ratio of the (square of the) turbulent to large-scale
magnetic field strength $\left\langle B_{\mathrm{t}}^{2}\right\rangle /\left\langle B_{0}^{2}\right\rangle $.
As was shown in Paper I, this dispersion function is readily evaluated
from polarization data and the aforementioned ratio could be used
to calculate, for example, the large-scale magnetic field strength
$\left\langle B_{0}^{2}\right\rangle ^{1/2}$ through the Chandrasekhar-Fermi
equation \citep{CF1953}. It should be clear, however, that in order
to precisely achieve such a goal we must find a way to determine other
parameters such as the turbulent correlation scales $\delta$ and
$\delta^{\prime}$, the effective cloud thickness $\Delta^{\prime}$,
and the ratio of the turbulent to large-scale polarized emission $\left\langle F_{\mathrm{t}}^{2}\right\rangle /\left\langle F_{0}^{2}\right\rangle $.

The polarized emission, in particular, is problematic. This is because
we do not have direct information on the weighting function $F\left(\mathbf{r},z\right)$
or its integrated counterpart

\begin{equation}
\overline{F}\left(\mathbf{r}\right)=\iint H\left(\mathbf{r}-\mathbf{a}\right)\left[\frac{1}{\Delta}\int_{0}^{\Delta}F\left(\mathbf{a},z\right)dz\right]d^{2}a.\label{eq:Fbar}\end{equation}

Although the integrated polarized flux $\overline{P}\left(\mathbf{r}\right)$
is contained in polarization data sets obtained with dust continuum
polarimeters such as Hertz \citep{Dotson2009}, SCUBA \citep{Matthews2009},
and SHARP \citep{Vaillancourt2008}, it does not correspond to the
quantity defined in Equation (\ref{eq:Fbar}). Instead the polarized
flux is observationally determined through the measurement of the
integrated Stokes parameters 

\begin{eqnarray}
\overline{Q}\left(\mathbf{r}\right) & = & \iint H\left(\mathbf{r}-\mathbf{a}\right)\left[\frac{1}{\Delta}\int_{0}^{\Delta}Q\left(\mathbf{a},z\right)dz\right]d^{2}a\label{eq:Qbar}\\
\overline{U}\left(\mathbf{r}\right) & = & \iint H\left(\mathbf{r}-\mathbf{a}\right)\left[\frac{1}{\Delta}\int_{0}^{\Delta}U\left(\mathbf{a},z\right)dz\right]d^{2}a\label{eq:Ubar}\end{eqnarray}

\noindent with

\begin{equation}
\overline{P}\left(\mathbf{r}\right)=\sqrt{\overline{Q}^{2}\left(\mathbf{r}\right)+\overline{U}^{2}\left(\mathbf{r}\right)}.\label{eq:Pbar}\end{equation}

Although these parameters will also exhibit large-scale and turbulent
components, they do not provide us with any means for disentangling
the turbulent contributions $\left\langle B_{\mathrm{t}}^{2}\right\rangle $
and $\left\langle F_{\mathrm{t}}^{2}\right\rangle $ due to the magnetic
field and the polarized emission they contain. We must then resort
to some approximation(s) if we are to make any progress.

Perhaps the most obvious difference between the two turbulent terms
on the right-hand side of Equation (\ref{eq:cos_sol2}) is that the
first one is of first order in the square of the turbulent to large-scale
magnetic field strength ratio while the other is of second order.
If we assume that such a ratio is a fraction of unity, then the second
order term can be neglected. This amounts to neglecting the contribution
of the turbulent polarized emission in Equation (\ref{eq:cos_sol2});
this is the line of attack we will use from now on. We therefore write
that

\begin{equation}
1-\left\langle \cos\left[\Delta\Phi\left(\ell\right)\right]\right\rangle \simeq\sqrt{2\pi}\frac{\left\langle B_{\mathrm{t}}^{2}\right\rangle }{\left\langle B_{0}^{2}\right\rangle }\left[\frac{\delta^{3}}{\left(\delta^{2}+2W^{2}\right)\Delta^{\prime}}\right]\left[1-e^{-\ell^{2}/2\left(\delta^{2}+2W^{2}\right)}\right]+\sum_{j=1}^{\infty}a_{2j}^{\prime}\ell^{2j},\label{eq:1-cos}\end{equation}

\noindent while for displacements $\ell$ less than a few times $W$
we keep only the first $\ell^{2}$ term in the Taylor expansion

\begin{equation}
1-\left\langle \cos\left[\Delta\Phi\left(\ell\right)\right]\right\rangle \simeq\sqrt{2\pi}\frac{\left\langle B_{\mathrm{t}}^{2}\right\rangle }{\left\langle B_{0}^{2}\right\rangle }\left[\frac{\delta^{3}}{\left(\delta^{2}+2W^{2}\right)\Delta^{\prime}}\right]\left[1-e^{-\ell^{2}/2\left(\delta^{2}+2W^{2}\right)}\right]+a_{2}^{\prime}\ell^{2},\label{eq:1-cos_app}\end{equation}

\noindent or

\begin{equation}
\left\langle \Delta\Phi^{2}\left(\ell\right)\right\rangle \simeq2\sqrt{2\pi}\frac{\left\langle B_{\mathrm{t}}^{2}\right\rangle }{\left\langle B_{0}^{2}\right\rangle }\left[\frac{\delta^{3}}{\left(\delta^{2}+2W^{2}\right)\Delta^{\prime}}\right]\left[1-e^{-\ell^{2}/2\left(\delta^{2}+2W^{2}\right)}\right]+m^{2}\ell^{2}.\label{eq:Delta_Phi2_app}\end{equation}

Whether it is appropriate or not to neglect the turbulent polarized
emission is an open question, which could perhaps be adequately investigated
through simulations.

\subsection{The Effective Cloud Thickness\label{sub:thickness}}

We previously defined the effective cloud thickness $\Delta^{\prime}$
through the following relation

\begin{eqnarray}
\left\langle \alpha\left(0\right)\right\rangle  & = & \iint\iint H\left(\mathbf{a}\right)H\left(\mathbf{a}^{\prime}\right)\left\{ \frac{2}{\Delta}\int_{0}^{\Delta}\left(1-\frac{u}{\Delta}\right)\left[\frac{R_{F,0}\left(v,u\right)}{\left\langle F_{0}^{2}\right\rangle }\right]\left[\frac{R_{\mathbf{B},0}\left(v,u\right)}{\left\langle B_{0}^{2}\right\rangle }\right]du\right\} d^{2}a^{\prime}d^{2}a\nonumber \\
 & \equiv & \frac{\Delta^{\prime}}{\Delta}\leq1.\label{eq:Delta'_again}\end{eqnarray}

\noindent We can get a sense of the nature of $\Delta^{\prime}$ by
expressing $\left\langle \alpha\left(0\right)\right\rangle $ with
the Fourier transform of $\left\langle \alpha\left(\ell\right)\right\rangle $
at $\ell=0$ (see Section \ref{sub:ls} in Appendix \ref{sec:Derivations}
for more details)

\begin{equation}
\left\langle \alpha\left(0\right)\right\rangle =\frac{1}{\left(2\pi\right)^{2}}\iint\left\Vert H\left(\mathbf{k}_{v}\right)\right\Vert ^{2}\left\{ \frac{1}{2\pi}\int\left[\frac{\mathcal{R}_{0}\left(\mathbf{k}_{v},k_{u}\right)}{\left\langle F_{0}^{2}\right\rangle \left\langle B_{0}^{2}\right\rangle }\right]\mathrm{sinc}^{2}\left(\frac{k_{u}\Delta}{2}\right)dk_{u}\right\} d^{2}k_{v},\label{eq:alpha(0)_FT}\end{equation}

\noindent where $\mathcal{R}_{0}\left(\mathbf{k}_{v},k_{u}\right)$
is the Fourier transform of $\mathcal{R}_{0}\left(v,u\right)=R_{F,0}\left(v,u\right)R_{\mathbf{B},0}\left(v,u\right)$
and $\mathrm{sinc}\left(x\right)\equiv\sin\left(x\right)/x$. It should
be apparent that Equation (\ref{eq:alpha(0)_FT}) includes several
effects that set the value of $\Delta^{\prime}$. For the idealized
case where the autocorrelation $\mathcal{R}_{0}\left(v,u\right)$
is uniform across the cloud, its Fourier transform $\mathcal{R}_{0}\left(\mathbf{k}_{v},k_{u}\right)$
is proportional to a Dirac distribution (more precisely, $\mathcal{R}_{0}\left(\mathbf{k}_{v},k_{u}\right)=\left(2\pi\right)^{3}\left\langle F_{0}^{2}\right\rangle \left\langle B_{0}^{2}\right\rangle \delta\left(\mathbf{k}_{v},k_{u}\right)$)
and $\Delta^{\prime}=\Delta$, as was previously stated. For more
realistic cases, however, $\mathcal{R}_{0}\left(\mathbf{k}_{v},k_{u}\right)$
will have a finite width along $k_{u}$ and $\mathbf{k}_{v}$. As
is also discussed in Section \ref{sub:ls}, the (square of) beam profile
$\left\Vert H\left(\mathbf{k}_{v}\right)\right\Vert ^{2}$ will filter
$\mathcal{R}_{0}\left(\mathbf{k}_{v},k_{u}\right)$ along $\mathbf{k}_{v}$
and therefore reduce the value of $\left\langle \alpha\left(0\right)\right\rangle $
(i.e., $\Delta^{\prime}<\Delta$) through the exclusion of spectral
modes located outside the bandwidth subtended by $\left\Vert H\left(\mathbf{k}_{v}\right)\right\Vert ^{2}$.
A more important filtering effect due to the finite spectral width
of $\mathcal{R}_{0}\left(\mathbf{k}_{v},k_{u}\right)$ along $k_{u}$
is expected, however, as it is severely truncated by the integration
process through the cloud thickness $\Delta$. This is clearly assessed
by the presence $\mathrm{sinc}^{2}\left(k_{u}\Delta/2\right)$ in
Equation (\ref{eq:alpha(0)_FT}) and the subsequent integration on
$k_{u}$. Because of the assumed large difference between the cloud
thickness and the beam radius (i.e., $W\ll\Delta$) we expect that
this effect will dominate in determining the value of $\left\langle \alpha\left(0\right)\right\rangle $. 

If we now introduce $\Delta k_{u}$ the spectral width of $\mathcal{R}_{0}\left(\mathbf{k}_{v},k_{u}\right)$
along $k_{u}$ with the following definition

\begin{equation}
2\Delta k_{u}\mathcal{R}_{0}\left(\mathbf{k}_{v},0\right)\equiv\int\mathcal{R}_{0}\left(\mathbf{k}_{v},k_{u}\right)dk_{u},\label{eq:dku}\end{equation}

\noindent then we can further approximate (since $\Delta k_{u}\gg\Delta^{-1}$)

\begin{eqnarray}
\int\mathcal{R}_{0}\left(\mathbf{k}_{v},k_{u}\right)\mathrm{sinc}^{2}\left(\frac{k_{u}\Delta}{2}\right)dk_{u} & \sim & \mathcal{R}_{0}\left(\mathbf{k}_{v},0\right)\int\mathrm{sinc}^{2}\left(\frac{k_{u}\Delta}{2}\right)dk_{u}\nonumber \\
 & \sim & \frac{2\pi}{\Delta}\mathcal{R}_{0}\left(\mathbf{k}_{v},0\right)\nonumber \\
 & \sim & \frac{2\pi}{\Delta}\left[\frac{1}{2\Delta k_{u}}\int\mathcal{R}_{0}\left(\mathbf{k}_{v},k_{u}\right)dk_{u}\right].\label{eq:dku_R_0}\end{eqnarray}

\noindent Inserting this last relation into Equation (\ref{eq:alpha(0)_FT}),
with the assumption that the filtering due to the telescope beam is
negligible compared to that due to integration through the thickness
of the cloud, yields

\begin{eqnarray}
\left\langle \alpha\left(0\right)\right\rangle  & \sim & \left(\frac{2\pi}{\Delta}\right)\left(\frac{1}{2\Delta k_{u}}\right)\left\{ \frac{1}{\left(2\pi\right)^{3}}\iiint\left[\frac{\mathcal{R}_{0}\left(\mathbf{k}_{v},k_{u}\right)}{\left\langle F_{0}^{2}\right\rangle \left\langle B_{0}^{2}\right\rangle }\right]d^{3}k\right\} \nonumber \\
 & \sim & \left(\frac{\pi}{\Delta}\right)\left(\frac{1}{\Delta k_{u}}\right)\nonumber \\
 & \equiv & \frac{\Delta^{\prime}}{\Delta}.\label{eq:Delta'_alpha(0)}\end{eqnarray}

\noindent We then find 

\begin{equation}
\Delta k_{u}\sim\frac{\pi}{\Delta^{\prime}}.\label{eq:dku_Delta'}\end{equation}

\noindent Because of the way the Fourier transform relates the width
of $\mathcal{R}_{0}\left(\mathbf{k}_{v},k_{u}\right)$ to that of
$\mathcal{R}_{0}\left(v,u\right)$, $\Delta^{\prime}$ can therefore
be advantageously interpreted as the width of the large-scale autocorrelation
function $\mathcal{R}_{0}\left(v,u\right)$ (if we assume isotropy,
then the width is the same along $u$ or $v$). 

Unfortunately, we do not have access to $\mathcal{R}_{0}\left(v,u\right)$
when mapping the polarization of dust emission in a molecular cloud.
However, we can probably get a decent approximation for $\Delta^{\prime}$
through the shape of the autocorrelation function of the aforementioned
cloud- and beam-integrated polarized flux $\overline{P}\left(\mathbf{r}\right)$.
This function is defined with

\begin{equation}
\left\langle \overline{P}^{2}\left(\ell\right)\right\rangle \equiv\left\langle \overline{P}\left(\mathbf{r}\right)\overline{P}\left(\mathbf{r}+\boldsymbol{\ell}\right)\right\rangle .\label{eq:auto_Pbar}\end{equation}
Albeit this is an integrated quantity, it does provide a good sense
of the proportion of the cloud that contains the bulk of the polarized
flux, which is used as the weighting function for the magnetic field
in our analysis (see Eq. {[}\ref{eq:Bbar}{]}). We also note that
this autocorrelation is a function of the distance $\ell$ on the
surface of the cloud, not through its depth. But this is consistent
with the isotropy assumption from which it is expected that a molecular
cloud will have similar characteristics through its depth and across
its surface. It therefore seems reasonable to associate the width
of $\left\langle \overline{P}^{2}\left(\ell\right)\right\rangle $
with the effective depth of the cloud. 

\begin{figure}
\epsscale{0.9}\plotone{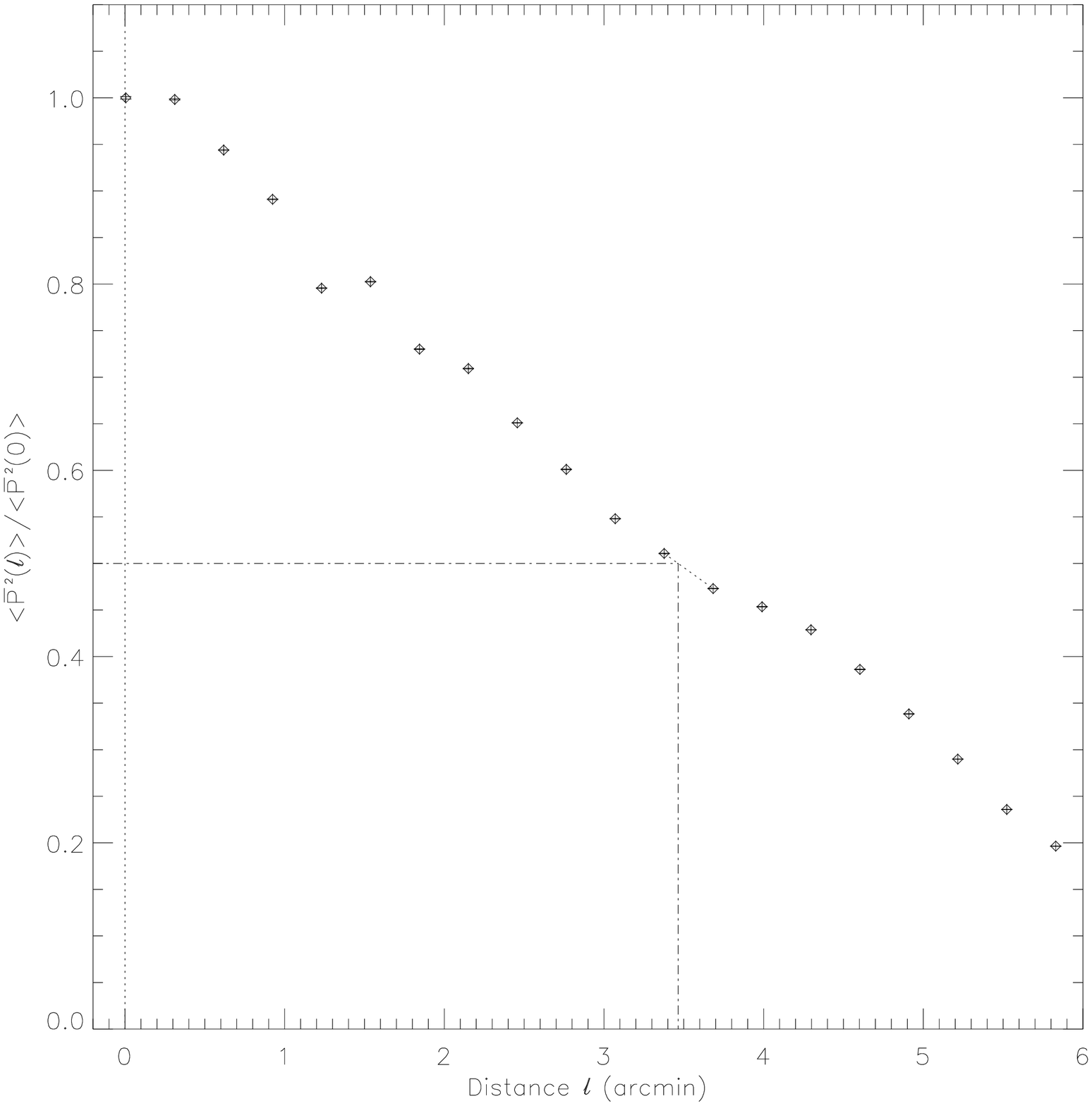}

\caption{\label{fig:corrpf}Normalized autocorrelation function of the integrated
polarized flux as calculated using the previously published 350-$\mu$m
Hertz data for OMC-1 \citep{Houde2004b, Hildebrand2009}. We chose
the width at half magnitude to determine the value for the effective
depth of OMC-1; we therefore have $\Delta^{\prime}\approx3\farcm5$.}

\end{figure}

For the case of OMC-1 we use previously published Hertz data%
\footnote{We use data obtained with Hertz instead of SHARP as the former cover
a larger spatial extent and allow us to determine $\left\langle \overline{P}^{2}\left(\ell\right)\right\rangle $
over a large enough distance.%
} \citep{Houde2004b, Hildebrand2009} to evaluate the effective depth
of the cloud. The result is presented in Figure \ref{fig:corrpf},
where the normalized autocorrelation function of the integrated normalized
flux is shown. As indicated on the graph, we have arbitrarily chosen
the width at half magnitude as the value for $\Delta^{\prime}$; we
will therefore set $\Delta^{\prime}\approx3\farcm5$ for OMC-1 for
the calculations that will follow. Also notably, we redefine $N$
as follows (see Eq. {[}\ref{eq:N}{]})

\begin{equation}
N=\frac{\left(\delta^{2}+2W^{2}\right)\Delta^{\prime}}{\sqrt{2\pi}\,\delta^{3}}.\label{eq:newN}\end{equation}

\subsection{The Turbulent Correlation Length Scale, the Turbulent to Large-scale
Magnetic Field Strength Ratio, and the Large-scale Magnetic Field
Strength\label{sub:fit} }

Having estimated the effective depth of the cloud $\Delta^{\prime}$
from the autocorrelation function of the integrated normalized flux,
we are now in a position to determine two fundamental parameters that
characterize magnetic fields and turbulence in star-forming regions:
the turbulent correlation length $\delta$ and the (square of the)
turbulent to large-scale magnetic field strength ratio $\left\langle B_{\mathrm{t}}^{2}\right\rangle /\left\langle B_{0}^{2}\right\rangle $.
To do so we refer to Equation (\ref{eq:1-cos_app}), which we write
again here for convenience

\begin{equation}
1-\left\langle \cos\left[\Delta\Phi\left(\ell\right)\right]\right\rangle \simeq\sqrt{2\pi}\frac{\left\langle B_{\mathrm{t}}^{2}\right\rangle }{\left\langle B_{0}^{2}\right\rangle }\left[\frac{\delta^{3}}{\left(\delta^{2}+2W^{2}\right)\Delta^{\prime}}\right]\left[1-e^{-\ell^{2}/2\left(\delta^{2}+2W^{2}\right)}\right]+a_{2}^{\prime}\ell^{2},\label{eq:1-cos_app2}\end{equation}

\noindent which is valid when the displacement $\ell$ is less than
a few times $W$.

Our plan consists of using the previously published 350-$\mu$m SHARP
polarization map of OMC-1 \citep{Vaillancourt2008} to evaluate the
left-hand side of Equation (\ref{eq:1-cos_app2}) and fit our solution
to the problem (i.e., the right-hand side) to the data. There are
only three quantities to be simultaneously fitted for: $\delta$,
$\left\langle B_{\mathrm{t}}^{2}\right\rangle /\left\langle B_{0}^{2}\right\rangle $,
and $a_{2}^{\prime}$, the first two being the parameters we are most
interested in at this time. More details concerning our data analysis
will be found in Appendix \ref{sec:Data}. 

We show in Figure \ref{fig:struct} the result of our non-linear fit
to the aforementioned data. In the figure, the top graph shows the
fit of Equation (\ref{eq:1-cos_app2}) (solid curve) to the data (symbols)
when plotted as a function of $\ell^{2}$. The broken curve does not
contain the correlated part of the function, i.e., the function 

\begin{equation}
\sqrt{2\pi}\frac{\left\langle B_{\mathrm{t}}^{2}\right\rangle }{\left\langle B_{0}^{2}\right\rangle }\left[\frac{\delta^{3}}{\left(\delta^{2}+2W^{2}\right)\Delta^{\prime}}\right]+a_{2}^{\prime}\ell^{2}\label{eq:m2+bl2}\end{equation}

\noindent is displayed to better visualize the integrated turbulent
contribution (i.e., the first term) to the dispersion in relation
to the large-scale component (i.e., $a_{2}^{\prime}\ell^{2}$) when
$\ell$ is less than a few times $W$. The middle graph of Figure
\ref{fig:struct} displays the same information as the top graph but
plotted as a function of $\ell$\emph{.} In the bottom graph we show
the correlated turbulent component of the dispersion function (symbols)
obtained by subtracting the data to the broken curve in the middle
graph, which in our model (solid curve) corresponds to 

\begin{eqnarray}
b^{2}\left(\ell\right) & \equiv & \frac{\left\langle \overline{\mathbf{B}}_{\mathrm{t}}\cdot\overline{\mathbf{B}}_{\mathrm{t}}\left(\ell\right)\right\rangle }{\left\langle B_{0}^{2}\right\rangle }\nonumber \\
 & = & \sqrt{2\pi}\frac{\left\langle B_{\mathrm{t}}^{2}\right\rangle }{\left\langle B_{0}^{2}\right\rangle }\left[\frac{\delta^{3}}{\left(\delta^{2}+2W^{2}\right)\Delta^{\prime}}\right]e^{-\ell^{2}/2\left(\delta^{2}+2W^{2}\right)}.\label{eq:turb_comp}\end{eqnarray}

Finally, the broken curve shows what would be the expected contribution
of the (assumed Gaussian) telescope beam alone to the width of the
turbulent component (i.e, when $\delta=0$ in the argument of the
exponential). Although we performed our analysis for a Gaussian turbulent
correlation function, it is rather unlikely that the autocorrelation
function of the turbulent magnetic field component fits this model.
We therefore did not use the first few points (i.e., for $\ell\lesssim0\farcm2$)
to fit Equation (\ref{eq:1-cos_app2}) to the data and concentrated
on larger values of $\ell$ where it is more likely to obtain a reasonable
fit. Indeed, there is evidence from the first few points, when $\ell\lesssim0\farcm2$,
in the middle and bottom graphs of Figure \ref{fig:struct} that the
Gaussian turbulent autocorrelation function assumption is incorrect,
as expected. 

A comparison of the expected contribution of the telescope beam to
the correlated turbulent component of the dispersion function (broken
curve in the bottom graph of Figure \ref{fig:struct}) with the accompanying
data reveals the imprint of the finite turbulent correlation length
scale $\delta$ in the relative excess detected in the $0\farcm1\lesssim\ell\lesssim0\farcm4$
range. This imprint is also evident from the non-zero integrated ratio
of the (square of the) turbulent to large-scale magnetic field strength
ratio (i.e., $b^{2}\left(0\right)$ from Equation {[}\ref{eq:turb_comp}{]})
that is evident in the data through the intercept of the broken curves
at $\ell=0$ in the top and middle graph of Figure \ref{fig:struct}.
As is seen in Equation (\ref{eq:turb_comp}) this intercept will go
to zero in the limit where $\delta\rightarrow0$ when the number of
turbulent cells subtended by the telescope beam $N$ tends to infinity
(see Equation {[}\ref{eq:newN}{]}), i.e., in cases where the turbulent
component is basically completely integrated out. 

The results from our fit of Equation (\ref{eq:1-cos_app2}) to the
dispersion data are summarized in Table \ref{tab:Results}. Most notably,
we measure the turbulent correlation length to be $\delta\approx16$
mpc (or $7\farcs3$ at 450 pc, the distance we adopt for OMC-1), which
implies that there are on average $N\approx21$ independent turbulent
cells contained within the column of gas probed by our telescope beam.
Furthermore, since our fit for the square of the {}``integrated''
turbulent to large-scale magnetic strength ratio yielded $b^{2}\left(0\right)\approx0.013$,
then it follows that 

\begin{eqnarray}
\frac{\left\langle B_{\mathrm{t}}^{2}\right\rangle }{\left\langle B_{0}^{2}\right\rangle } & = & Nb^{2}\left(0\right)\nonumber \\
 & \approx & 0.28,\label{eq:ratio2}\end{eqnarray}

\noindent where we used Equations (\ref{eq:newN}) and (\ref{eq:turb_comp}).
This quantity can be inserted in the Chandrasekhar-Fermi equation
\citep{CF1953} to evaluate the strength plane-of-the-sky component
of the large-scale magnetic field

\begin{eqnarray}
\left\langle B_{0}^{2}\right\rangle ^{1/2} & = & \sqrt{4\pi\rho}\,\sigma\left(v\right)\left[\frac{\left\langle B_{\mathrm{t}}^{2}\right\rangle }{\left\langle B_{0}^{2}\right\rangle }\right]^{-1/2}\nonumber \\
 & \approx & 760\,\mu\mathrm{G},\label{eq:CF}\end{eqnarray}

\noindent where we used the same values as in Paper I for the mass
density $\rho$ (i.e., a gas density of $10^{5}$ cm$^{-3}$ and a
mean molecular weight of 2.3) and the one-dimensional velocity dispersion
$\sigma\left(v\right)$ (i.e., $1.85\,\mathrm{km}\,\mathrm{s}^{-1}$
as obtained from a representative H$^{13}$CO $J=3\rightarrow2$ spectrum). 

We are now in a better position to appreciate the importance of adequately
taking into account the signal integration process. Indeed, failing
to do so while using the Chandrasekhar-Fermi equation would imply
multiplying the value obtained in Equation (\ref{eq:CF}) by $\sqrt{N}\approx4.6$.
This would yield a magnetic field strength of approximately 3.5 mG,
a value that is certainly prohibitively high for OMC-1 (see Section
\ref{sub:spectrum}). 

\begin{deluxetable}{ccccccc}

\tablewidth{0pt}
\tablecolumns{7}

\tablehead{
\multicolumn{3}{c}{Fit Result} & \colhead{} & \multicolumn{3}{c}{Derived Quantities} \\

\cline{1-3} \cline{5-7} \\

\colhead{$\delta$\tablenotemark{a}} & \colhead{$b^{2}\left(0\right)$\tablenotemark{b}} & \colhead{$a_{2}^{'}$} & \colhead{} 
& \colhead{$N$\tablenotemark{c}} & \colhead{$\left\langle B_{\mathrm{t}}^{2}\right\rangle /\left\langle B_{0}^{2}\right\rangle$\tablenotemark{d}} & \colhead{$\left\langle B_{0}^2\right\rangle^{1/2}$\tablenotemark{e}} \\

\colhead{(mpc)} & \colhead{} & \colhead{($\mathrm{arcmin}^{-2}$)} & \colhead{} & \colhead{} & \colhead{} & \colhead{($\mu\mathrm{G}$)}
}

\tablecaption{Results from our fit of Equation (\ref{eq:1-cos_app2}) to the dispersion data for OMC-1.  \label{tab:Results}}

\startdata

$16.0\pm0.4$ & $0.0134\pm0.001$ & $0.059\pm0.001$ & & 20.7 & $0.28\pm0.01$ & 760

\enddata

\tablenotetext{a}{Corresponds to the fit result of $7\farcs3\pm0\farcs2$ at the distance of 450 pc assumed for OMC-1.}
\tablenotetext{b}{Corresponds to the linear intercept of the broken curves at $\ell=0$ in the top and middle graphs of Figure \ref{fig:struct} (see Equation [\ref{eq:turb_comp}]).}  
\tablenotetext{c}{Calculated using Equation (\ref{eq:newN}) with the SHARP beam radius $W=4\farcs7$ (or $\mathrm{FWHM}=11^{''}$) and $\Delta^{'}=3\farcm5$.}
\tablenotetext{d}{Calculated by multiplying the fit result for $b^{2}\left(0\right)$ by $N$ (see Equations [\ref{eq:newN}] and [\ref{eq:turb_comp}]).}
\tablenotetext{e}{Calculated using the Chandrasekhar-Fermi equation (see Equation [\ref{eq:CF}]) assuming a density of $10^5$ cm$^{-3}$, a mean molecular weight of 2.3, and a velocity dispersion of $1.85 \mathrm{\ km\ s}^{-1}$. This estimate is probably not precise to better than a factor of a few due to the uncertainty in the density and $\Delta^{'}$.}

\end{deluxetable}

\section{Discussion\label{sec:Discussion}}

\subsection{The Turbulent Power Spectrum\label{sub:spectrum}}

The determination of the turbulent correlation $\delta$ length is
important for the characterization of turbulence in molecular clouds.
It is therefore desirable to compare our result of $\delta\approx16$
mpc for OMC-1 with other independent techniques or analyses that seek
to evaluate this quantity, or others related to it, either observationally
or theoretically.

A parameter that is very closely related to $\delta$ is the ambipolar
diffusion scale $\delta_{\mathrm{AD}}$ at which the ionized and neutral
components of the gas decouple, and that may determine the cutoff
wavelength of the power spectrum of the turbulent component of the
magnetic field $\mathbf{B}_{\mathrm{t}}$. For this discussion we
therefore define $k_{\mathrm{AD}}=\delta_{\mathrm{AD}}^{-1}$ such
that the (isotropic) turbulent power spectrum $\mathcal{R}_{\mathrm{t}}\left(k\right)\approx0$
for $k>k_{\mathrm{AD}}$. It follows that if $\delta^{-1}$ is some
measure of the width of $\mathcal{R}_{\mathrm{t}}\left(k\right)$
(e.g., its standard deviation), then we expect that $\delta\geq\delta_{\mathrm{AD}}$. 

Recent theoretical \citep{Lazarian2004} and observational studies
have yielded estimates of the order of 1 mpc for $\delta_{\mathrm{AD}}$
in molecular clouds. Although an observational determination for the
ambipolar diffusion scale has yet to be obtained for OMC-1, \citet{Li2008b}
have measured $\delta_{\mathrm{AD}}\approx2$ mpc for the molecular
cloud M17. If we assume for the moment that this value also applies
to OMC-1, then we find that the turbulent correlation length scale
of approximately 16 mpc we measured, which is a factor of many larger
than the quoted ambipolar diffusion cutoff scale, is consistent with
the aforementioned expectation that $\delta\geq\delta_{\mathrm{AD}}$.

One should always keep in mind, however, that the value for $\delta_{\mathrm{AD}}$
determined by \citet{Li2008b} pertains to a {}``Kolmogorov-like''
turbulent power spectrum whereas our value for $\delta$ does not.
That is, given a three-dimensional isotropic turbulent autocorrelation
function%
\footnote{For this discussion we do not restrict $\ell$ to the plane of the
sky, but allow it to span the three dimensions.%
} $\mathcal{R}_{\mathrm{t}}\left(\ell\right)$ from which one evaluates
the correlation length $\delta$, the corresponding power spectrum
$\mathcal{R}_{\mathrm{t}}\left(k\right)$ (of width $\delta^{-1}$)
is specified by the Fourier transform of $\mathcal{R}_{\mathrm{t}}\left(\ell\right)$.
The related Kolmogorov-like power spectrum is not $\mathcal{R}_{\mathrm{t}}\left(k\right)$,
however, but is usually defined as $\mathcal{R}_{\mathrm{K}}\left(k\right)\equiv4\pi k^{2}\mathcal{R}_{\mathrm{t}}\left(k\right)$
\citep{Frisch1995}. It follows that if $\delta_{\mathrm{AD}}$ is
the ambipolar diffusion scale pertaining to the Kolmogorov-like spectrum
$\mathcal{R}_{\mathrm{K}}\left(k\right)$, then we should define a
turbulent correlation length $\delta_{\mathrm{K}}$ that also pertains
to $\mathcal{R}_{\mathrm{K}}\left(k\right)$ for a meaningful comparison.

For example, if we consider the idealization of the Gaussian turbulent
autocorrelation function of width $\delta$ used for our analysis 

\begin{equation}
\mathcal{R}_{\mathrm{t}}\left(\ell\right)=e^{-\ell^{2}/2\delta^{2}},\label{eq:R_t(ell)}\end{equation}

\noindent then 

\begin{equation}
\mathcal{R}_{\mathrm{t}}\left(k\right)=\left(2\pi\right)^{3/2}\delta^{3}e^{-\frac{1}{2}\delta^{2}k^{2}}\label{eq:R_t(k)}\end{equation}

\noindent and the Kolmogorov-like power spectrum is given by

\begin{equation}
\mathcal{R}_{\mathrm{K}}\left(k\right)=2\left(2\pi\right)^{5/2}\delta^{3}k^{2}e^{-\frac{1}{2}\delta^{2}k^{2}}.\label{eq:R_K(k)}\end{equation}

Calculating the (square of the) width $\delta_{\mathrm{K}}^{-1}$
of $\mathcal{R}_{\mathrm{K}}\left(k\right)$ (i.e., its variance)
we have

\begin{eqnarray}
\delta_{\mathrm{K}}^{-2} & = & \frac{\int_{0}^{\infty}k^{2}\mathcal{R}_{\mathrm{K}}\left(k\right)dk}{\int_{0}^{\infty}\mathcal{R}_{\mathrm{K}}\left(k\right)dk}\nonumber \\
 & = & 3\delta^{-2}.\label{eq:delta_K}\end{eqnarray}

Applying this relation to the result obtained for OMC-1, we now find
that $\delta_{\mathrm{K}}\approx9$ mpc. This value is now a factor
of almost two closer to theoretically and observationally expected
values for $\delta_{\mathrm{AD}}$. Furthermore, we should also keep
in mind that our assumption of Gaussian functions is probably incorrect.
The true turbulent autocorrelation function could yield different
values for $\delta$ and $\delta_{\mathrm{K}}$, which may be closer
to the expected value for $\delta_{\mathrm{AD}}$.

Another fundamental parameter for the characterization of turbulence
is the turbulent to large-scale magnetic field strength ratio. With
the expectation that the magnetic field will be tied to the gas through
flux-freezing, which should apply for a significant part of the spectrum
where $k<\delta_{\mathrm{AD}}^{-1}$, this parameter is a measure
of the relative amount of turbulent energy contained in the gas. Our
aforementioned determined value of 0.28 for the square of the turbulent
to large-scale magnetic field strength ratio seems to indicate that
turbulence does not dominate the dynamics in OMC-1.

As is shown in Equation (\ref{eq:CF}), the square root of this ratio
is used to calculate the strength plane-of-the-sky component of the
large-scale magnetic field with

\begin{equation}
\left[\frac{\left\langle B_{\mathrm{t}}^{2}\right\rangle }{\left\langle B_{0}^{2}\right\rangle }\right]^{1/2}\approx0.53.\label{eq:sqr_ratio}\end{equation}

\noindent Although our estimate for the plane-of-the-sky component
of the large-scale magnetic field of $\left\langle B_{0}^{2}\right\rangle ^{1/2}\approx760\,\mu\mathrm{G}$
calculated using this ratio cannot be precise to better than a factor
of a few because of uncertainties in the gas density and $\Delta^{\prime}$,
this value is reasonable and in line with other independent measurements.
For example, \citet{Crutcher1999} measured a line-of-sight magnetic
field strength of $360\,\mu\mathrm{G}$ in OMC-1 using CN Zeeman measurements,
which probed densities that are comparable to those corresponding
to our observations. Incidentally, using these values for the two
components of the magnetic field we can get an estimate of approximately
$65^{\circ}$ for the inclination angle of the large scale magnetic
field to the line of sight. Again, this value is consistent with the
results obtained by \citet{Houde2004b} for this object using an independent
technique (i.e., $49^{\circ}$ in the Orion bar and $65^{\circ}$
at a location a few arcminutes northeast of Orion KL).

\subsection{Weaknesses of the Technique}

The use of the dispersion function in the polarization angle, while
adequately taking into account the process of signal integration implicit
to dust polarization measurements, allowed us to determine some of
the fundamental parameters characterizing magnetized turbulence in
molecular clouds. Our analysis rests, however, on a few assumptions
that require some discussion:
\begin{itemize}
\item Our definition for the integrated magnetic field $\overline{\mathbf{B}}\left(\mathbf{r}\right)$
(see Eq. {[}\ref{eq:Bbar}{]}) does not perfectly mimic the measurement
process through which dust polarization data are obtained. A consequence
of this was noted in the discussion that followed Equation (\ref{eq:cos_sol1}),
where it was found that our analysis seemingly introduces a contribution
to the angular dispersion that we do not expect to be present in actual
data. Although a hypothetical analysis that would precisely duplicate
the measurement process (see Eqs. {[}\ref{eq:Qbar}{]}-{[}\ref{eq:Pbar}{]})
would be desirable, the quest for an analytical solution renders this
kind of idealization necessary. The same comment could be made for
the assumptions of isotropy, homogeneity, and stationarity used throughout
our calculations. Nonetheless, we expect that the results stemming
from our treatment of the angular dispersion function are successful
in the characterization of magnetized turbulence in molecular clouds. 
\item We assumed a Gaussian form for the autocorrelation functions characterizing
turbulence. This assumption is certainly incorrect and as a result
it was not possible to obtain a reasonable fit when using part of
the data where $\ell\lesssim0\farcm2$; the shortcomings of our Gaussian
model is most clearly apparent in that region (see the discussion
in Section \ref{sub:fit}). We must therefore keep in mind that our
estimates for the turbulent correlation length $\delta$ and the turbulent
to large-scale magnetic field strength ratio are correspondingly uncertain
to some extent. On the other hand our fit is relatively robust, as
small changes in the domain used for the fit (i.e., the range of values
for $\ell$) do not lead to significantly different solutions. The
same is true if a term proportional to $\ell^{4}$ is added to the
large-scale function to be fitted (i.e., to the right-hand side of
Eq. {[}\ref{eq:1-cos_app2}{]}).
\item Another source of uncertainty is our modeling of the telescope beam
using a Gaussian profile. Although the beam size we quote is based
on actual measurements taken from chopped measurements on pointing
sources (e.g., Uranus) and is consistent with other SHARP observations,
we do not possess a detailed map of the telescope beam profile for
our set of observations. The aforementioned uncertainty is a consequence
of the fact that our determination of $\delta$ stems, in part, from
a comparison of the correlated turbulent component of the dispersion
function with the assumed Gaussian telescope beam (see the bottom
graph of Fig. \ref{fig:struct}). The significant difference observed
between the two, however, makes it unlikely that the ensuing effect
on our estimate of the turbulent length scale is important.
\item Our analysis has shown that the presence of turbulence in the polarized
emission can bring about additional angular dispersion to that due
to turbulence in the magnetic field. As was discussed in Section \ref{sub:pol},
however, our inability to disentangle these two contributions to the
dispersion has forced us to introduce approximations that essentially
brought about the neglect of the turbulent polarized emission. It
should be noted that this implies that we overestimated the amount
of turbulence in the magnetic field (measured through $\left\langle B_{\mathrm{t}}^{2}\right\rangle /\left\langle B_{0}^{2}\right\rangle $,
as obtained through our fit to the data), which in turn translates
into an underestimate of the large-scale magnetic strength when using
the Chandrasekhar-Fermi equation. Although this effect is probably
small, we once again stress that the effect of turbulence in the polarized
emission could perhaps be advantageously investigated and quantified
through simulations.
\end{itemize}
It is interesting to note that although we limited ourselves to the
determination of only two  parameters characterizing the turbulent
power spectrum, our technique can in principle be used to achieve
significantly more. We can verify this statement by considering the
Fourier transform $b^{2}\left(k_{v}\right)$ associated with the turbulent
component (see Eq. {[}\ref{eq:turb_comp}{]}, for example), which
can be determined from Equation (\ref{eq:autoBbar(kv)})

\begin{equation}
b^{2}\left(k_{v}\right)=\frac{1}{\left\langle B_{0}^{2}\right\rangle }\left\Vert H\left(\mathbf{k}_{v}\right)\right\Vert ^{2}\left[\int\mathcal{R}_{\mathrm{t}}\left(\mathbf{k}_{v},k_{u}\right)\mathrm{sinc}^{2}\left(\frac{k_{u}\Delta}{2}\right)dk_{u}\right],\label{eq:b2(k)}\end{equation}

\noindent where $\mathcal{R}_{\mathrm{t}}\left(\mathbf{k}_{v},k_{u}\right)$
is the Fourier transform of $\mathcal{R}_{\mathrm{t}}\left(v,u\right)\equiv R_{F,\mathrm{t}}\left(v,u\right)R_{\mathbf{B,\mathrm{t}}}\left(v,u\right)$.
Since the spatial frequency component $k_{u}$ of $\mathcal{R}_{\mathrm{t}}\left(\mathbf{k}_{v},k_{u}\right)$
is eliminated through the corresponding integration, we can equally
write

\begin{equation}
b^{2}\left(k_{v}\right)=\frac{1}{\left\langle B_{0}^{2}\right\rangle }\left\Vert H\left(\mathbf{k}_{v}\right)\right\Vert ^{2}\mathcal{R}_{\mathrm{t}}\left(\mathbf{k}_{v}\right),\label{eq:newb2(k)}\end{equation}

\noindent where $\mathcal{R}_{\mathrm{t}}\left(\mathbf{k}_{v}\right)$
is now interpreted as the two-dimensional turbulent power spectrum.
Under the assumption of isotropy it would be expected that $\mathcal{R}_{\mathrm{t}}\left(\mathbf{k}_{v}\right)$
is similar in form to $\mathcal{R}_{\mathrm{t}}\left(k\right)$, where
$\mathcal{R}_{\mathrm{t}}\left(k\right)$ is the three-dimensional
turbulent power spectrum defined in Section \ref{sub:spectrum}. Since
the left-hand side of Equation (\ref{eq:newb2(k)}) can be evaluated
by taking a Fourier transform of the data (as shown with the symbols
in the bottom graph of Fig. \ref{fig:struct}) and that the beam profile
$H\left(\mathbf{k}_{v}\right)$ is presumably well characterized,
it follows that the turbulent power spectrum $\mathcal{R}_{\mathrm{t}}\left(\mathbf{k}_{v}\right)$
can readily be determined through the inversion of Equation (\ref{eq:newb2(k)})
(using a simple Wiener optimal filter, for example). 

Unfortunately our polarization map does not have enough spatial resolution
to allow us to perform this analysis (i.e., our beam profile $H\left(\mathbf{k}_{v}\right)$
is much too narrow in frequency space), but there is no obvious reason
why this should not be feasible with higher resolution observations.
A much better characterization of the turbulent power spectrum would
then result. We will address this question in a subsequent publication.

\section{Summary\label{sec:Summary}}

In this paper we expanded our study on the dispersion of polarization
angles in molecular clouds and showed how the effect of signal integration
through the thickness of the cloud as well as across the area subtended
by the telescope beam inherent to dust continuum measurements can
be incorporated in our analysis. We correctly accounted for its effect
on the measured angular dispersion and inferred turbulent to large-scale
magnetic field strength ratio. We also showed how to evaluate the
turbulent magnetic field correlation scale from polarization data
and applied our results to the molecular cloud OMC-1. For this object,
we find a turbulent correlation length of $\delta\approx16$ mpc,
a turbulent to large-scale magnetic field strength ratio of approximately
0.5, and a plane-of-the-sky large-scale magnetic field strength $\left\langle B_{0}^{2}\right\rangle ^{-1/2}\approx760\,\mu\mathrm{G}$.

In future papers we will extend our technique to study the possibility
for obtaining a detailed characterization of the turbulent power spectrum,
as well as the discussing evidence for anisotropy in magnetized turbulence
from dust polarization data.

\acknowledgements{M.H.'s research is funded through the NSERC Discovery Grant, Canada
Research Chair, Canada Foundation for Innovation, Ontario Innovation
Trust, and Western's Academic Development Fund programs. J.E.V. acknowledges
support from the CSO, which is funded through NSF AST 05-40882. This
work has also been supported in part by NSF grants AST 05-05230, AST
02-41356, and AST 05-05124.}

\appendix
\section{Derivations}\label{sec:Derivations}

\subsection{The Autocorrelation Function of the Integrated Magnetic Field\label{sub:auto}}

\noindent To obtain Equation (\ref{eq:auto_Bbar}) we start with

\small

\begin{equation}
\left\langle \overline{\mathbf{B}}\left(\mathbf{r}\right)\cdot\overline{\mathbf{B}}\left(\mathbf{r}+\boldsymbol{\ell}\right)\right\rangle =\iint\iint H\left(\mathbf{r}-\mathbf{a}\right)H\left(\mathbf{r}+\boldsymbol{\ell}-\mathbf{a}^{\prime}\right)\left[\frac{1}{\Delta^{2}}\int_{0}^{\Delta}\int_{0}^{\Delta}R_{F}\left(v,u\right)R_{\mathbf{B}}\left(v,u\right)dz^{\prime}dz\right]d^{2}a^{\prime}d^{2}a\label{eq:autoBbar_a}\end{equation}

\normalsize

\noindent from Equations (\ref{eq:Bbar}), (\ref{eq:autoB}), (\ref{eq:auto_g}),
$u=\left|z^{\prime}-z\right|$, and $v=\left|\mathbf{a}^{\prime}-\mathbf{a}\right|$.
The integral

\begin{equation}
I\left(v\right)=\frac{1}{\Delta^{2}}\int_{0}^{\Delta}\int_{0}^{\Delta}R_{F}\left(v,u\right)R_{\mathbf{B}}\left(v,u\right)dz^{\prime}dz\label{eq:I(v)_double}\end{equation}

\noindent can be transformed into a one-dimensional integral on account
of the assumed stationarity of the functions present in the integrand
(i.e., they are solely a function of $\left|z^{\prime}-z\right|$,
besides $v$). To accomplish this, after making the change of variable
$u=\left|z^{\prime}-z\right|$, we integrate over the square surface
delimited by $0\leq z\leq\Delta$ and $0\leq z^{\prime}\leq\Delta$
along the linear path where $-\Delta\leq u\leq\Delta$. This path
is also perpendicular to a family of strips (along which $u$ is constant)
of infinitesimal area 

\begin{equation}
dS=\Delta\left(1-\frac{\left|u\right|}{\Delta}\right)du\label{eq:dS}\end{equation}

\noindent to first order in $du$. We then find that 

\begin{equation}
I\left(v\right)=\frac{2}{\Delta}\int_{0}^{\Delta}\left(1-\frac{u}{\Delta}\right)R_{F}\left(v,u\right)R_{\mathbf{B}}\left(v,u\right)du.\label{eq:I(v)}\end{equation}

We can now further make the change of variables $\mathbf{a}\rightarrow\mathbf{r}-\mathbf{a}$
and $\mathbf{a}^{\prime}\rightarrow\mathbf{r}-\mathbf{a}^{\prime}$,
which is then inserted with Equation (\ref{eq:I(v)_double}) into
Equation (\ref{eq:autoBbar_a}) to yield

\begin{equation}
\left\langle \overline{\mathbf{B}}\mathbf{\cdot}\overline{\mathbf{B}}\mathbf{\left(\boldsymbol{\ell}\right)}\right\rangle =\iint\iint H\left(\mathbf{a}\right)H\left(\mathbf{a}^{\prime}+\boldsymbol{\ell}\right)\left[\frac{2}{\Delta}\int_{0}^{\Delta}\left(1-\frac{u}{\Delta}\right)R_{F}\left(v,u\right)R_{\mathbf{B}}\left(v,u\right)du\right]d^{2}a^{\prime}d^{2}a,\label{eq:autoBbar_app}\end{equation}

\noindent where any dependency on $\mathbf{r}$ is done away with
since $u$ and $v$ are unaffected by this change of variables and
we integrate over all of space. 

Equation (\ref{eq:auto_Bbar_sol}) can be derived by first noting
that 

\begin{equation}
\frac{2}{\Delta}\int_{0}^{\Delta}\left(1-\frac{u}{\Delta}\right)e^{-\left(v^{2}+u^{2}\right)/2\delta^{2}}du\simeq\sqrt{2\pi}\left(\frac{\delta}{\Delta}\right)e^{-v^{2}/2\delta^{2}}\label{eq:onedim_integ}\end{equation}

\noindent when $\delta\ll\Delta$. This result will arise for all
terms containing a turbulent component (i.e., $\left\langle B_{\mathrm{t}}^{2}\right\rangle $
and/or $\left\langle F_{\mathrm{t}}^{2}\right\rangle $) in Equation
(\ref{eq:autoBbar_app}).

Setting $\mathbf{u}=\mathbf{a}^{\prime}+\mathbf{a}$ and $\mathbf{v}=\mathbf{a}^{\prime}-\mathbf{a}$
we have 

\begin{eqnarray}
\frac{1}{2W^{2}}\left(\left|\mathbf{a}\right|^{2}+\left|\mathbf{a}^{\prime}+\boldsymbol{\ell}\right|^{2}\right)+\frac{1}{2\delta^{2}}\left|\mathbf{a}^{\prime}-\mathbf{a}\right|^{2} & = & \frac{1}{4W^{2}}\left(\left|\mathbf{u}+\boldsymbol{\ell}\right|^{2}+\left|\mathbf{v}+\boldsymbol{\ell}\right|^{2}\right)+\frac{\left|\mathbf{v}\right|^{2}}{2\delta^{2}}\nonumber \\
 & = & \frac{\left|\mathbf{u}+\boldsymbol{\ell}\right|^{2}}{4W^{2}}+\left(\frac{\delta^{2}+2W^{2}}{4\delta^{2}W^{2}}\right)\left|\mathbf{v}+\boldsymbol{\ell}\left(\frac{\delta^{2}}{\delta^{2}+2W^{2}}\right)\right|^{2}\nonumber \\
 &  & +\frac{\ell^{2}}{2\left(\delta^{2}+2W^{2}\right)}.\label{eq:exponent}\end{eqnarray}

\noindent It follows from this relation, the Jacobian related to the
coordinate transformation above, and Equation (\ref{eq:beam}) that

\small

\begin{eqnarray}
\iint\iint H\left(\mathbf{a}\right)H\left(\mathbf{a}^{\prime}+\boldsymbol{\ell}\right)e^{-\left|\mathbf{a}^{\prime}-\mathbf{a}\right|^{2}/2\delta^{2}}d^{2}a^{\prime}d^{2}a & = & \frac{e^{-\ell^{2}/2\left(\delta^{2}+2W^{2}\right)}}{4\left(2\pi W^{2}\right)^{2}}\left[\int e^{-u^{2}/4W^{2}}du\cdot\int e^{-\left(\frac{\delta^{2}+2W^{2}}{4\delta^{2}W^{2}}\right)v^{2}}dv\right]^{2}\nonumber \\
 & = & \left(\frac{\delta^{2}}{\delta^{2}+2W^{2}}\right)e^{-\ell^{2}/2\left(\delta^{2}+2W^{2}\right)},\label{eq:beam_integ}\end{eqnarray}

\normalsize

\noindent which is dependent on only $\ell$ and not its orientation.
Equation (\ref{eq:auto_Bbar_sol}) follows from Equations (\ref{eq:auto_Bbar}),
(\ref{eq:autoB_gen}), (\ref{eq:autog_gen}), (\ref{eq:B_0^2})-(\ref{eq:norm_ls}),
(\ref{eq:onedim_integ}), and (\ref{eq:beam_integ}).

\subsection{The Length Scale of the Large-scale Component\label{sub:ls}}

We have used the region where the distance $\ell$ is less than a
few times the beam radius to fit the dispersion data to Equation (\ref{eq:1-cos_app}).
It is therefore appropriate to inquire as to the validity of this
procedure. To do so, we first note that the autocorrelation function
given by Equation (\ref{eq:auto_Bbar}) can be expanded with its Fourier
transform as follows (with $\mathcal{R}\left(v,u\right)\equiv R_{F}\left(v,u\right)R_{\mathbf{B}}\left(v,u\right)$)

\begin{eqnarray}
\left\langle \overline{\mathbf{B}}\left(\mathbf{r}\right)\cdot\overline{\mathbf{B}}\left(\mathbf{r}+\boldsymbol{\ell}\right)\right\rangle  & = & \iint\iint H\left(\mathbf{a}\right)H\left(\mathbf{a}^{\prime}+\boldsymbol{\ell}\right)\left[\frac{1}{\Delta}\int_{-\Delta}^{\Delta}\left(1-\frac{u}{\Delta}\right)\mathcal{R}\left(v,u\right)du\right]d^{2}a^{\prime}d^{2}a\nonumber \\
 & = & \iint\iint H\left(\mathbf{a}\right)H\left(\mathbf{a}^{\prime}+\boldsymbol{\ell}\right)\nonumber \\
 &  & \,\,\,\,\,\,\,\,\,\,\,\,\,\,\,\,\,\,\,\,\left\{ \frac{1}{\Delta}\int_{-\Delta}^{\Delta}\left(1-\frac{\left|u\right|}{\Delta}\right)\left[\frac{1}{\left(2\pi\right)^{3}}\iiint\mathcal{R}\left(\mathbf{k}_{v},k_{u}\right)e^{i\mathbf{k}\cdot\mathbf{x}}d^{3}k\right]du\right\} d^{2}a^{\prime}d^{2}a\nonumber \\
 & = & \frac{1}{\left(2\pi\right)^{3}}\iiint\mathcal{R}\left(\mathbf{k}_{v},k_{u}\right)\left[\iint\iint H\left(\mathbf{a}\right)H\left(\mathbf{a}^{\prime}+\boldsymbol{\ell}\right)e^{i\mathbf{k}_{v}\cdot\left(\mathbf{a}-\mathbf{a}^{\prime}\right)}d^{2}a^{\prime}d^{2}a\right]\nonumber \\
 &  & \,\,\,\,\,\,\,\,\,\,\,\,\,\,\,\,\,\,\,\,\,\,\,\,\,\,\,\,\,\left[\frac{1}{\Delta}\int_{-\Delta}^{\Delta}\left(1-\frac{\left|u\right|}{\Delta}\right)e^{ik_{u}u}du\right]d^{3}k\nonumber \\
 & = & \frac{1}{\left(2\pi\right)^{3}}\iiint\mathcal{R}\left(\mathbf{k}_{v},k_{u}\right)\left[H\left(\mathbf{k}_{v}\right)H\left(-\mathbf{k}_{v}\right)e^{i\mathbf{k}_{v}\cdot\boldsymbol{\ell}}\right]\nonumber \\
 &  & \,\,\,\,\,\,\,\,\,\,\,\,\,\,\,\,\,\,\,\,\,\,\,\,\,\,\,\,\,\left[\frac{1}{\Delta}\int_{-\Delta}^{\Delta}\left(1-\frac{\left|u\right|}{\Delta}\right)e^{ik_{u}u}du\right]d^{3}k,\label{eq:autoBbar(k)}\end{eqnarray}

\noindent where the Fourier transform of a function is represented
by simply replacing the spatial arguments by their $\mathbf{k}$-space
counterparts (e.g., $\mathcal{R}\left(v,u\right)\rightleftharpoons\mathcal{R}\left(\mathbf{k}_{v},k_{u}\right)$),
and $\mathbf{x}=\mathbf{v}+u\mathbf{e}_{z}$ with $\mathbf{v}=\mathbf{a}^{\prime}-\mathbf{a}$
and $u=\left|z^{\prime}-z\right|$. But since

\begin{eqnarray}
\frac{1}{\Delta}\int_{-\Delta}^{\Delta}\left(1-\frac{\left|u\right|}{\Delta}\right)e^{ik_{u}u}du & = & \frac{1}{\Delta^{2}}\int\left[\int\mathrm{rect}\left(\frac{\tau}{\Delta}\right)\mathrm{rect}\left(\frac{u-\tau}{\Delta}\right)d\tau\right]e^{ik_{u}u}du\nonumber \\
 & = & \mathrm{sinc}^{2}\left(\frac{k_{u}\Delta}{2}\right),\label{eq:sinc2}\end{eqnarray}

\noindent with

\begin{equation}
\mathrm{rect}\left(\frac{\tau}{\Delta}\right)=\left\{ \begin{array}{cc}
1, & \left|\tau\right|<\frac{\Delta}{2},\\
0, & \left|\tau\right|>\frac{\Delta}{2},\end{array}\right.\label{eq:rect}\end{equation}

\noindent we can write 

\begin{equation}
\left\langle \overline{\mathbf{B}}\left(\mathbf{r}\right)\cdot\overline{\mathbf{B}}\left(\mathbf{r}+\boldsymbol{\ell}\right)\right\rangle =\frac{1}{\left(2\pi\right)^{2}}\iint\left\Vert H\left(\mathbf{k}_{v}\right)\right\Vert ^{2}\left[\frac{1}{2\pi}\int\mathcal{R}\left(\mathbf{k}_{v},k_{u}\right)\mathrm{sinc}^{2}\left(\frac{k_{u}\Delta}{2}\right)dk_{u}\right]e^{i\mathbf{k}_{v}\cdot\boldsymbol{\ell}}d^{2}k_{v},\label{eq:autoBbar(kv)}\end{equation}

\noindent where $\left\Vert H\left(\mathbf{k}_{v}\right)\right\Vert ^{2}=H\left(\mathbf{k}_{v}\right)H^{*}\left(\mathbf{k}_{v}\right)$
and $\mathrm{sinc}\left(x\right)\equiv\sin\left(x\right)/x$.

The effect of signal integration is made explicit in Equation (\ref{eq:autoBbar(kv)}).
More precisely, it is seen that the integration through the thickness
of the cloud heavily filters the spectral components along the line
of sight to a small set symmetrically located about $k_{u}=0$ through
the presence of $\mathrm{sinc}^{2}\left(k_{u}\Delta\right)$ and the
integration along $k_{u}$, while the integration across the beam
profile on the plane of the sky further filters the signal with $\left\Vert H\left(\mathbf{k}_{v}\right)\right\Vert ^{2}$.
For a Gaussian profile as specified by Equation (\ref{eq:beam}) we
have

\begin{equation}
H\left(\mathbf{k}_{v}\right)=e^{-\frac{1}{2}W^{2}k_{v}^{2}},\label{eq:H(kv)}\end{equation}

\noindent and it becomes clear the only spectral components that remain
in the integrated polarization map satisfy $k_{v}\lesssim W^{-1}$;
the larger the telescope beam the more heavily the signal is filtered
spatially.

It should also be noted that the derivation of Equation (\ref{eq:auto_Bbar_sol}),
which was demonstrated in Section \ref{sub:auto}, can also be achieved
using Equation (\ref{eq:autoBbar(kv)}) under the assumption of Gaussian
turbulent autocorrelation functions (see Equations {[}\ref{eq:autoB_gen}{]}
and {[}\ref{eq:autog_gen}{]}) with $\delta\ll\Delta$.

Restricting Equation (\ref{eq:autoBbar(kv)}) to its large-scale component,
assuming isotropy, and referring to Equation (\ref{eq:norm_ls}) we
can write

\begin{eqnarray}
\left\langle \alpha\left(\ell\right)\right\rangle  & = & \frac{1}{\left(2\pi\right)^{2}}\int_{0}^{2\pi}\int_{0}^{\infty}\left\Vert H\left(k_{v}\right)\right\Vert ^{2}\left[\frac{1}{2\pi}\int\frac{\mathcal{R}_{0}\left(k_{v},k_{u}\right)}{\left\langle F_{0}^{2}\right\rangle \left\langle B_{0}^{2}\right\rangle }\mathrm{sinc}^{2}\left(\frac{k_{u}\Delta}{2}\right)dk_{u}\right]e^{ik_{v}\ell\cos\left(\phi\right)}k_{v}dk_{v}d\phi\nonumber \\
 & = & \frac{1}{2\pi}\int_{0}^{\infty}\left\Vert H\left(k_{v}\right)\right\Vert ^{2}J_{0}\left(k_{v}\ell\right)\left[\frac{1}{2\pi}\int\frac{\mathcal{R}_{0}\left(k_{v},k_{u}\right)}{\left\langle F_{0}^{2}\right\rangle \left\langle B_{0}^{2}\right\rangle }\mathrm{sinc}^{2}\left(\frac{k_{u}\Delta}{2}\right)dk_{u}\right]k_{v}dk_{v}.\label{eq:alpha_FT-1}\end{eqnarray}

\noindent where $\mathcal{R}_{0}\left(v,u\right)=R_{F,0}\left(v,u\right)R_{\mathbf{B},0}\left(v,u\right)$
and $J_{0}\left(x\right)$ is the Bessel function of the first kind
of order $0$. Performing a Taylor expansion about $\ell=0$ using
the identity

\begin{equation}
J_{0}\left(x\right)=\sum_{n=0}^{\infty}\left(-1\right)^{n}\frac{\left(x^{2}/4\right)^{n}}{\left(n!\right)^{2}}\label{eq:J0}\end{equation}

\noindent it is clear that, as expected, a Taylor expansion of $\left\langle \alpha\left(\ell\right)\right\rangle $
only contains terms of even power. Equations (\ref{eq:H(kv)}) and
(\ref{eq:alpha_FT-1}) show that, once again, the beam filtering process
will remove any spectral component with $k_{v}\gtrsim W^{-1}$ in
the large-scale dispersion function; its characteristic length scale
is therefore constrained by, and limited to a few times, the telescope
beam radius. Because of this beam filtering effect we expect the part
of the spectrum at $k_{v}\lesssim W^{-1}$ to be dominant in Equation
(\ref{eq:alpha_FT-1}), therefore limiting ourselves to a domain where
$\ell$ is less than a few times the beam radius (i.e., $k_{v}\ell\lesssim1$)
to fit the dispersion data with Equation (\ref{eq:1-cos_app}) implies
that

\begin{equation}
J_{0}\left(k_{v}\ell\right)\simeq1-\frac{\left(k_{v}\ell\right)^{2}}{4},\label{eq:J0_app}\end{equation}

\noindent which justifies the method used to model that data and extract
the needed information from it.

\section{Data Analysis}\label{sec:Data}

The data for OMC-1 from the SHARP polarimeter studied here have been
previously published by \citet{Vaillancourt2008}. For our purposes
we only include data that satisfy the $p>3\sigma_{p}$ criterion,
where $p$ is the polarization fraction and $\sigma_{p}$ its uncertainty.

The angle differences between each and every pair of data points are
calculated as

\begin{equation}
\Delta\Phi_{ij}=\Phi_{i}-\Phi_{j},\label{eq:diff}\end{equation}
and the corresponding distance between each point

\begin{equation}
\ell_{ij}\equiv\vert\mathbf{r}_{i}-\mathbf{r}_{j}\vert.\label{eq:distance}\end{equation}

\noindent Note that $\ell_{ij}=\ell_{ji}$ so that a map with $N$
data points contains only $N(N-1)/2$ distinct differences. Also note
that $\left|\Delta\Phi_{ij}\right|$ is constrained to be in the range
$[0,90]$ degrees.

These data are divided into separate distance bins with sizes corresponding
to integer multiples of the grid spacing $\Delta\ell$ that results
after processing the SHARP map ($\Delta\ell=2\farcs37$; note that
a SHARP pixel is approximately $4\farcs6$ and the beam FWHM for our
polarization map is approximately $11^{\prime\prime}$); the bin for
$\ell_{k}$ (which corresponds to $k$ pixels) covers $\left(\ell_{k}-\Delta\ell/2\right)\leq\ell_{ij}<\left(\ell_{k}+\Delta\ell/2\right)$.
Within each bin $k$ we thus calculate the dispersion function with

\begin{equation}
1-\left\langle \cos\left(\Delta\Phi_{ij}\right)\right\rangle _{k},\,\,\,\,\,\mathrm{for\, all\,\,\,\,}\left(\ell_{k}-\Delta\ell/2\right)\leq\ell_{ij}<\left(\ell_{k}+\Delta\ell/2\right).\label{eq:rms}\end{equation}

The dispersion function is corrected for measurement uncertainty within
each bin according to the uncertainty on each $\Delta\Phi_{ij}$ and
propagating the measurement uncertainties on both $\Phi_{i}$ and
$\Phi_{j}$ available in the data set, as is explained below. However,
since it is often the case that $\ell_{ij}<W$ some values for $\Delta\Phi_{ij}$
and their corresponding uncertainty will be correlated. We have therefore
used the following relation for the measurement uncertainty

\begin{equation}
\sigma^{2}(\Delta\Phi_{ij})\simeq\sigma^{2}(\Phi_{i})+\sigma^{2}(\Phi_{j})-2\sigma(\Phi_{i})\sigma(\Phi_{j})e^{-\ell_{ij}^{2}/4W^{2}},\label{eq:sigma2_dphi}\end{equation}

\noindent which exhibits the right behavior when $\ell_{ij}=0$ and
$\ell_{ij}\gtrsim2W$. The data are then corrected for measurement
uncertainty with 

\begin{eqnarray}
\left\langle \cos\left(\Delta\Phi_{ij}\right)\right\rangle _{k,0} & = & \frac{\left\langle \cos\left(\Delta\Phi_{ij}\right)\right\rangle _{k}}{\left\langle \cos\left[\sigma\left(\Delta\Phi_{ij}\right)\right]\right\rangle _{k}}\nonumber \\
 & \simeq & \frac{\left\langle \cos\left(\Delta\Phi_{ij}\right)\right\rangle _{k}}{1-\frac{1}{2}\left\langle \sigma^{2}(\Delta\Phi_{ij})\right\rangle _{k}},\label{eq:cos(Dphi)_app}\end{eqnarray}

\noindent where $\left\langle \cos\left(\Delta\Phi_{ij}\right)\right\rangle _{k,0}$
is the quantity used in our analysis and for the plots in Figure \ref{fig:struct},
and an even probability distribution was assumed for the measurement
uncertainty in $\Delta\Phi_{ij}$.

Finally, the measurement uncertainties for the dispersion function
$1-\left\langle \cos\left(\Delta\Phi_{ij}\right)\right\rangle _{k,0}$
is determined with 

\begin{equation}
\sigma^{2}\left[\left\langle \cos\left(\Delta\Phi_{ij}\right)\right\rangle _{k,0}\right]=\left\langle \sin\left(\Delta\Phi_{ij}\right)\right\rangle _{k}^{2}\left\langle \sigma^{2}(\Delta\Phi_{ij})\right\rangle _{k}+\frac{3}{4}\left\langle \cos\left(\Delta\Phi_{ij}\right)\right\rangle _{k}^{2}\left\langle \sigma^{4}(\Delta\Phi_{ij})\right\rangle _{k},\label{eq:sigma2(cos)}\end{equation}

\noindent for all $\left(\ell_{k}-\Delta\ell/2\right)\leq\ell_{ij}<\left(\ell_{k}+\Delta\ell/2\right)$.
These uncertainties are those plotted in Figure \ref{fig:struct},
most of them are too small to be seen in the figure, especially at
the smallest displacements. 

\section{List of Symbols}\label{sec:Symbols}

In this appendix we list and define symbols for important variables
and functions appearing in the text, as well as give in parentheses
the number of the equation where they are first introduced or defined.
\begin{itemize}
\item $\mathbf{b}$: normalized magnetic field (Eq. {[}\ref{eq:b(r,z)}{]})
\item $\overline{\mathbf{b}}$: integrated normalized magnetic field (Eq.
{[}\ref{eq:bbar}{]})
\item $b^{2}\left(\ell\right)$: correlated turbulent component (Eq. {[}\ref{eq:turb_comp}{]})
\item $\mathbf{B}$: total magnetic field (Eq. {[}\ref{eq:Btot}{]})
\item $\mathbf{B}_{0}$: large-scale magnetic field (Eq. {[}\ref{eq:Btot}{]})
\item $\mathbf{B}_{\mathrm{t}}$: turbulent magnetic field (Eq. {[}\ref{eq:Btot}{]})
\item $\overline{\mathbf{B}}$: integrated total magnetic field (Eq. {[}\ref{eq:Bbar}{]})
\item $\overline{\mathbf{B}}_{0}$: integrated large-scale magnetic field
(Eq. {[}\ref{eq:Bbar}{]})
\item $\overline{\mathbf{B}}_{\mathrm{t}}$: integrated turbulent magnetic
field (Eq. {[}\ref{eq:Bbar}{]})
\item $\mathbf{e}_{r}$: unit vector along $\mathbf{r}$ on the plane-of-the-sky
(Eq. {[}\ref{eq:x}{]})
\item $\mathbf{e}_{z}$: unit vector along the line-of-sight within a molecular
cloud (Eq. {[}\ref{eq:x}{]})
\item $F$: total polarized emission (Eq. {[}\ref{eq:bbar}{]})
\item $F_{0}$: large-scale polarized emission (Eq. {[}\ref{eq:g_decomp}{]})
\item $F_{\mathrm{t}}$: turbulent polarized emission (Eq. {[}\ref{eq:g_decomp}{]})in 
\item $\overline{F}$: integrated total polarized emission (Eq. {[}\ref{eq:Fbar}{]})
\item $H$: telescope beam profile (Eq. {[}\ref{eq:bbar}{]})
\item $J_{0}$: Bessel function of the first kind of order 0 (Eq. {[}\ref{eq:J0}{]})
\item $\boldsymbol{\ell}$: distance between two measurement positions (Eq.
{[}\ref{eq:cos_noint}{]})
\item $N$: number of independent turbulent cells in the gas column probed
by the beam (Eqs. {[}\ref{eq:N}{]} and {[}\ref{eq:newN}{]})
\item $P$: polarized flux (Eq. {[}\ref{eq:Pbar}{]})
\item $\overline{P}$: integrated polarized flux (Eq. {[}\ref{eq:Pbar}{]})
\item $Q$: Stokes parameter (Eq. {[}\ref{eq:Qbar}{]})
\item $\overline{Q}$: integrated Stokes parameter (Eq. {[}\ref{eq:Qbar}{]})
\item $\mathbf{r}$: position vector on the plane-of-the-sky (Eq. {[}\ref{eq:x}{]})
\item $R_{\mathbf{B}}$: autocorrelation function of the total magnetic
field (Eq. {[}\ref{eq:autoB}{]})
\item $R_{\mathbf{B},0}$: autocorrelation function of the large-scale magnetic
field (Eq. {[}\ref{eq:autoB_j}{]})
\item $R_{\mathbf{B},\mathrm{t}}$: autocorrelation function of the turbulent
magnetic field (Eq. {[}\ref{eq:autoB_j}{]})
\item $R_{F}$: autocorrelation function of the total polarized emission
(Eq. {[}\ref{eq:auto_g}{]})
\item $R_{F,0}$: autocorrelation function of the large-scale polarized
emission (Eq. {[}\ref{eq:autog_j}{]})
\item $R_{F,\mathrm{t}}$: autocorrelation function of the turbulent polarized
emission (Eq. {[}\ref{eq:autog_j}{]})
\item $\mathcal{R}$: product of $R_{F}$ and $R_{\mathbf{B}}$ (in normal
space; Eq. {[}\ref{eq:autoBbar(k)}{]})
\item $\mathcal{R}_{0}$: product of $R_{F,0}$ and $R_{\mathbf{B},0}$
(in normal space; Eq. {[}\ref{eq:alpha(0)_FT}{]})
\item $\mathcal{R}_{\mathrm{K}}\left(k\right)$: Kolmogorov-like power spectrum
(Eq. {[}\ref{eq:R_K(k)}{]})
\item $\mathcal{R}_{\mathrm{t}}$: turbulent correlation function (Eq. {[}\ref{eq:R_t(ell)}{]})
\item $U$: Stokes parameter (Eq. {[}\ref{eq:Ubar}{]})
\item $\overline{U}$: integrated Stokes parameter (Eq. {[}\ref{eq:Ubar}{]})
\item $W$: telescope beam radius (Eq. {[}\ref{eq:beam}{]})
\item $\mathbf{x}$: three-dimensional position vector (Eq. {[}\ref{eq:x}{]})
\item $z$: position along the line-of-sight within a molecular cloud (Eq.
{[}\ref{eq:x}{]})
\item $\alpha\left(\ell\right)$: normalized large-scale function (Eq. {[}\ref{eq:norm_ls}{]})
\item $\delta$: turbulent correlation length scale of the magnetic field
(Eq. {[}\ref{eq:autoB_gen}{]})
\item $\delta^{\prime}$: turbulent correlation length scale of the polarized
emission (Eq. {[}\ref{eq:autog_gen}{]})
\item $\delta_{\mathrm{K}}^{-1}$: Kolomogorov spectral width (Eq. {[}\ref{eq:delta_K}{]})
\item $\Delta$: maximum depth of a molecular cloud along the line-of-sight
(Eq. {[}\ref{eq:bbar}{]})
\item $\Delta^{\prime}$: effective depth of the molecular cloud along the
line-of-sight (Eq. {[}\ref{eq:Delta'}{]})
\item $\Delta k_{u}$: spectral width of $\mathcal{R}_{0}$ (Eq. {[}\ref{eq:dku}{]})
\item $\Delta\Phi\left(\boldsymbol{\ell}\right)$: difference in polarization
angles between two positions separated by $\boldsymbol{\ell}$ (Eq.
{[}\ref{eq:cos_noint}{]})
\item $\left\langle \cdots\right\rangle $: average of some quantity (Eq,
{[}\ref{eq:cos_noint}{]})\end{itemize}

\begin{figure}
\plotone{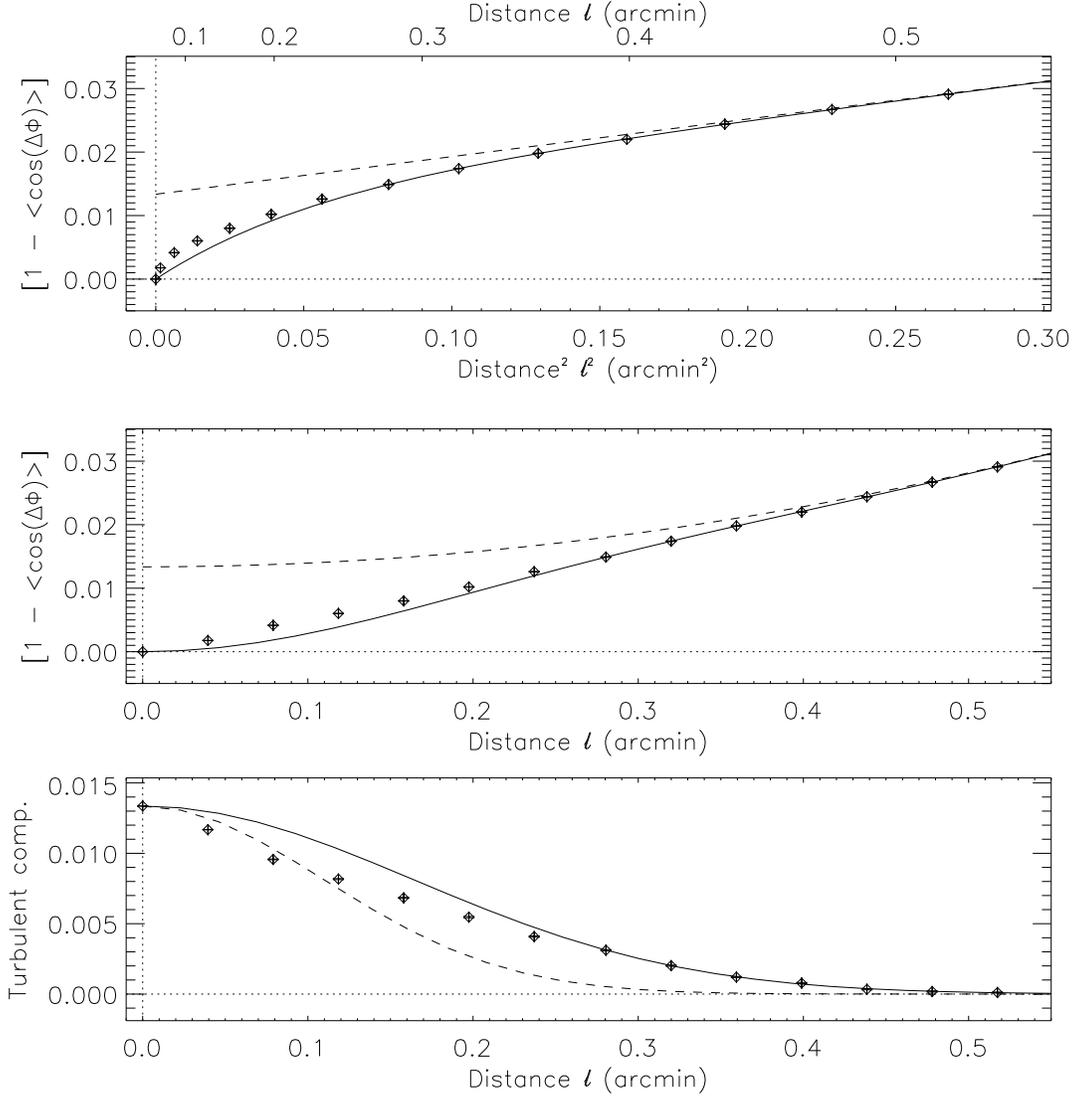}

\caption{\label{fig:struct}The dispersion function $1-\left\langle \cos\left[\Delta\Phi\left(\ell\right)\right]\right\rangle $
for OMC-1 using the 350-$\mu$m data obtained with SHARP. \emph{Top:}
fit of Equation (\ref{eq:1-cos_app}) (solid curve) to the data (symbols)
when plotted as a function of $\ell^{2}$, the broken curve does not
contain the correlated part of the function (see text); \emph{middle:}
same as top but plotted as a function of $\ell$; \emph{bottom:} the
turbulent component of the dispersion function (symbols), as obtained
by subtracting the data points to the broken curve in the middle graph,
while the broken and solid curves are, respectively, the contribution
of the (assumed Gaussian) telescope beam alone (i.e, when $\delta=0$)
and the fit to the data (i.e., with $\delta=7\farcs3$). }

\end{figure}


\begin{thebibliography}{Wiebe \& Watson(2004)}
\bibitem[Chandrasekhar \& Fermi(1953)]{CF1953}Chandrasekhar, S.,
and Fermi, E. 1953, \apj, 118, 113

\bibitem[Crutcher et al.(1999)]{Crutcher1999}Crutcher, R. M., Troland,
T. H., Lazareff, B., Paubert, G., and Kazès, I. 1999, \apj, L121

\bibitem[Dotson et al.(2009)]{Dotson2009}  Dotson, J. L., Davidson,
J. A., Dowell, C. D., Hildebrand, R. H., Kirby, L., and Vaillancourt,
J. E. 2009, submitted to \apjs

\bibitem[Dowell et al.(1998)]{Dowell1998}Dowell, C. D., Hildebrand,
R. H., Schleuning, D. A., Vaillancourt, J. E., Dotson, J. L., Novak,
G., Renbarger, T., and Houde, M. 1998, \apj, 504, 588

\bibitem[Falceta-Gonçalves et al.(2008)]{Falceta2008}Falceta-Gonçalves,
D., Lazarian, A., and Kowal, G. 2008, \apj, 679, 537

\bibitem[Frisch(1995)]{Frisch1995}Frisch, U. 1995, Turbulence: The
Legacy of A. N. Kolmogorov (Cambridge: Cambridge University Press)

\bibitem[Heitsch et al.(2001)]{Heitsch2001}Heitsch, F., Zweibel,
E. G., Mac Low, M-M., Li, P., and Norman, M. L. 2001, \apj, 561,
800

\bibitem[Hildebrand et al.(2009)]{Hildebrand2009}Hildebrand, R. H.,
Kirby, L. Dotson, J. L., Houde, M., and Vaillancourt, J. E. 2009,
\apj, 696, 567 (Paper I)

\bibitem[Houde et al.(2004)]{Houde2004b}Houde, M., Dowell, C. D.,
Hildebrand, R. H., Dotson, J. L., Vaillancourt, J. E., Phillips, T.
G., Peng, R., and Bastien, P. 2004, \apj, 604, 717

\bibitem[Kudoh \& Basu(2003)]{Kudoh2003}Kudoh, T., and Basu, S. 2003,
\apj, 595, 842

\bibitem[Lazarian et al.(2004)]{Lazarian2004}Lazarian, A., Vishniac,
E. T., and Cho, J. 2004, \apj, 603, 180

\bibitem[Li et al.(2008)]{Li2008a}Li, H., Dowell, C. D., Kirby, L.,
Novak, G., and Vaillancourt, J. E. 2008, \ao, 47, 422

\bibitem[Li \& Houde(2008)]{Li2008b}Li, H., and Houde, M. 2008, \apj,
677, 1151

\bibitem[Li et al.(2006)]{Li2006}Li, H., Attard, M., Dowell, C. D.,
Hildebrand. R. H., Houde, M., Kirby, L., Novak, G., and Vaillancourt,
J. E. 2006, \procspie, 6275, 48

\bibitem[Matthews et al.(2009)]{Matthews2009}Matthews, B. C., McPhee,
C., Fissel, L., \& Curran, R. L. 2009, \apjs submitted

\bibitem[Myers \& Goodman(1991)]{Myers1991}Myers, P. C., and Goodman,
A. A. 1991, \apj, 373, 509

\bibitem[Novak et al.(2004)]{Novak2004}Novak, G., et al. 2004, \procspie,
5498, 278

\bibitem[Ostriker et al.(2001)]{Ostriker2001}Ostriker, E. C., Stone,
J. M., and Gammie, C. F. 2001, \apj, 546, 980

\bibitem[Padoan et al.(2001)]{Padoan2001}Padoan, P., Goodman, A.
A., Draine, B. T., Juvela, M., Nordland, \AA ., and Rögnvaldsson,
Ö. E. 2001, \apj, 559, 1005

\bibitem[Vaillancourt et al.(2008)]{Vaillancourt2008}Vaillancourt,
J. E., Dowell, C. D., Hildebrand, R. H., Kirby, L., Krejny, M. M.,
Li, H., Novak, G., Houde, M., Shinnaga, H., and Attard, M. 2008, \apj,
675, L25

\bibitem[Wiebe \& Watson(2004)]{Wiebe2004}Wiebe, D. S., and Watson,
W. D. 2004, \apj, 615, 314

\end{thebibliography}
\end{document}